\documentclass[11pt,aps,prb,amsmath,amssymb,superscriptaddress,showpacs,floatfix,longbibliography]{revtex4-2}
\usepackage{graphicx} 
\usepackage{multirow}
\usepackage[table,xcdraw]{xcolor}
\usepackage{booktabs} 
\usepackage{amsmath}
\usepackage{xfrac}

\usepackage[hidelinks]{hyperref}
\usepackage{xr-hyper}  

\newcommand{\code}[1]{\texttt{#1}}
\newcommand{\bs}{\mathbf}

\def\INCLUDED{}

\begin{document}
\title{Benchmark of First-Principles Titanium K-Edge X-Ray Absorption Spectral Simulations on Titanium-containing Oxides}
\author{Chuntian Cao}
\affiliation{Computing and Data Sciences Directorate, Brookhaven National Laboratory, Upton, New York 11973, USA
}%

\author{Joshua J. Kas}
\affiliation{Department of Physics, University of Washington, Seattle, Washington 98195, United States
}%

\author{Karol Dyro}
\affiliation{Department of Physics and Astronomy, Stony Brook University, Stony Brook, New York 11794, United States
}%

\author{Bruce Ravel}
\email{bruce.ravel@nist.gov, 0000-0002-4126-872X}
\affiliation{Material Measurement Laboratory, National Institute of Standards and Technology, Gaithersburg, Maryland 20899, United States
}%

\author{John Vinson}
\email{john.vinson@nist.gov, orcid: 0000-0002-7619-7060}
\affiliation{Material Measurement Laboratory, National Institute of Standards and Technology, Gaithersburg, Maryland 20899, United States
}%

\author{Deyu Lu}
\email{dlu@bnl.gov, orcid: 0000-0003-4351-6085}
\affiliation{ 
Center for Functional Nanomaterials,
Brookhaven National Laboratory, Upton, New York 11973,
United States
}
\date{\today}
\begin{abstract}
X-ray absorption spectroscopy (XAS) is a powerful, element-specific probe for investigating the local structural and electronic properties of materials. However, quantitative analysis remains challenging, necessitating accurate first-principles spectral simulations. In this study, we benchmark first-principles simulations of Ti K-edge X-ray absorption near-edge structure (XANES) against experimental data for nine common titanium compounds. We systematically investigate key physical effects, including quadrupole excitations, thermal disorder, and many-body shake-up. Our results demonstrate that quadrupole excitations and thermal disorder are essential for capturing accurate pre-edge features, while many-body shake-up effects significantly influence the spectral shape of the main- and post-edge regions. By incorporating these effects alongside a band-theory treatment of the core-hole final state, our simulated spectra achieve excellent agreement with experimental data for most of the systems, as evidenced by high similarity scores. The shoulder peak in BaTiO$_3$ at 4980 eV in the experiment is largely missing in simulation. Further analysis shows that more accurate electronic structure theory than semi-local density functional theory is required to capture the correlation effects of the Ba $4f$ orbitals and that defects, such as oxygen vacancies, may also contribute to the shoulder. Beyond tackling specific titanium material systems, this work establishes a robust workflow for generating high-fidelity Ti K-edge XANES databases for titanium compounds, providing a framework that can be generalized to a broad range of materials.

\end{abstract}
\maketitle

\section{Introduction}

X-ray absorption near-edge structure (XANES) spectroscopy is a powerful technique for probing the local chemical environments of specific elements in materials during physical and chemical processes~\cite{RevModPhys.72.621}. 
First-principles simulations of XANES have become essential for interpreting experimental spectra and guiding materials discovery~\cite{Lu_Wang_JACS,Cao_Lu_ZnS,cao2024atomic,balugani24}. Improving the fidelity of simulated XANES spectra not only enhances the predictive power of the theory, but also enables data-driven workflows in materials science and strengthens the robustness of machine learning (ML) models trained on synthetic data. 

3d transition metals are present in material families with broad technological applications, including electronics, catalysis, energy storage, and spintronics. Their K-edge XANES spectra encodes crucial information about oxidation states and the coordination chemistry, offering physical insights into the structural, electronic and magnetic  properties~\cite{PhysRevB.76.214117,GLATZEL200565,van2017situ,yoon2005investigation,cabaret2010first,de2001high}.
However, the accurate simulation of K-edge XANES for 3\textit{d} transition metals is challenging, as multiple key physical effects need to be accounted for properly. The core-hole final-state effects are important for obtaining the right spectral shape. The pre-edge features are complex and arise from a mixture of quadrupole transitions and dipole transitions enabled by atomic \textit{p}-\textit{d} mixing or metal -- ligand -- neighboring metal hybridization~\cite{Yamamoto_review_preEdge}. 
The intensities of pre-edge peaks are sensitive to local inversion symmetry breaking from static distortion or thermal vibrations~\cite{cabaret2010first}. Many-body effects of the core-hole spectral function, such as shake-up satellites, can affect XANES spectral shapes~\cite{Calandra,JW1,JW2,Kas_PHC}. Neglecting these effects in XANES simulations will fail to reproduce the experimentally measured spectra. 

Previously, some of us and collaborators have carried out a multicode Ti K-edge XANES benchmark~\cite{MengFanchen_PhysRevMaterials} using three popular codes, {\sc ocean}~\cite{OCEAN2022, OCEAN_PhysRevB.83.115106}, \texttt{exciting}~\cite{gulans2014exciting,Vorwerk2019} and {\sc xspectra}~\cite{taillefumier2002x,gougoussis2009first}. In that study, the core-hole final-state effects are treated either by solving the Bethe-Salpeter equation (BSE) under the many-body perturbation theory formalism ({\sc ocean} and {\texttt {exciting}}) or the excited-electron core-hole method ({\sc xspectra}) where valence electrons are relaxed self-consistently in the presence of the core hole approximated by a core-hole pseudopotential. We established a rigorous procedure to achieve the numerical convergence in spectral simulations. The BSE codes ({\sc ocean} and {\texttt {exciting}}) and the core-hole potential code ({\sc xspectra}) show overall good agreement with each other, with some visible differences in the shoulder and main-edge spectral shape that can be primarily attributed to the difference in the strength of the screened core-hole potential~\cite{MengFanchen_PhysRevMaterials}. However, when compared to measured spectra of rutile and anatase TiO$_2$, simulations show several important discrepancies, including 1) the first pre-edge peak in rutile is significantly underestimated in simulation, 
2) the simulated intensity of the post-edge in both rutile and anatase decays too fast compared to experiment
and 3) in contrast to the experiment, the relative intensity of the rutile main edge doublet is inverted in {\sc ocean}. These discrepancies motivated us to carry out a follow-up study to benchmark simulation against experiment and systematically investigate the effects of thermal disorder and shake-up satellite, which are not considered in the previous work.

In this study, we benchmark \textit{ab initio} simulations of Ti K-edge XANES against measured data, focusing on the spectral range of about 40 eV, including pre-edge, main-edge, and post-edge features. We examine nine common Ti compounds, starting from experimental crystal structures, and incorporate three physical effects in addition to the band theory treatment of the core-hole final state effects (BSE and the core-hole pseudopotential method): (i) quadrupole transitions, (ii) thermal fluctuations, and (iii) many-body shakeup corrections. Our results show that quadrupole and thermal disorder effects primarily influence the pre-edge region, while many-body corrections are essential for capturing main-edge and post-edge shapes. By evaluating each individual effect, we assess their quantitative impact on spectral features. As a result, our study establishes a rigorous workflow for capturing key physics in Ti K-edge XANES simulations, which can play an important role in building high-quality simulation databases for XAS ML model development.


\section{Simulation workflow}
\begin{figure}[bthp]
        \centering
        \includegraphics[width=16cm]{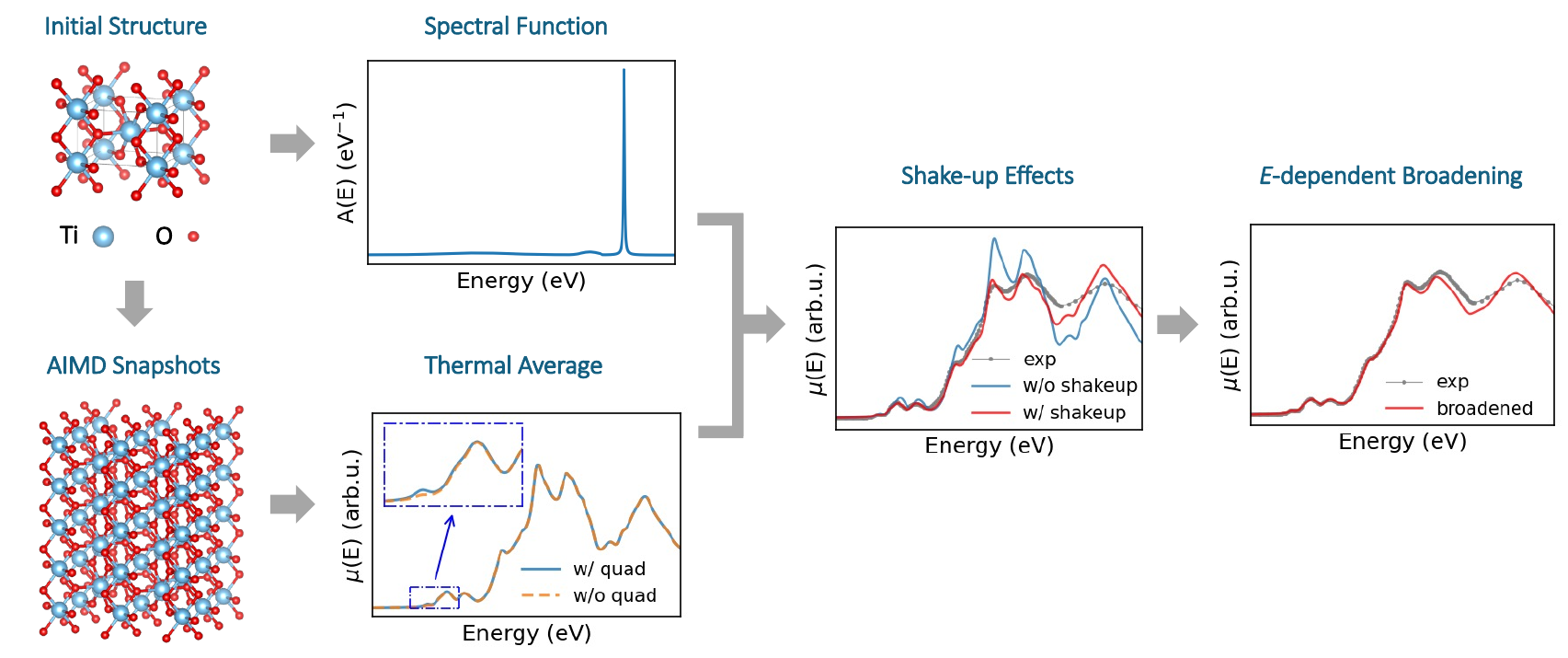}
        \vspace{-0mm}
        \caption{Schematic workflow of Ti K-edge XANES simulations.}
        \label{fig:workflow}
    \end{figure}

A robust XANES simulation workflow is essential for deciphering the complex structure -- spectrum relationship in spectral analysis. In the context of data-driven XANES analysis, such a workflow is the foundation of high-throughput simulations used to generate high-fidelity synthetic XANES spectra databases for downstream ML applications.  An overview of our XANES simulation workflow is shown in Fig.~\ref{fig:workflow}. This workflow takes into account key physical effects in first-principles XANES simulations and validates our resulting Ti K-edge XANES with measured experimental standards.

For each system in our material library, we extract the experimental crystal structure at room temperature (300 K) from crystallographic databases. The thermal disorder effects can be captured using the statistical average of the spectra, and a common practice is to average over snapshots from MD simulations. In addition, there are several other options. Cockayne \emph{et al.}~\cite{Cockayne_PhysRevB} extracted the magnitude of inversion-symmetry breaking phonon modes from  \emph{ab initio} molecular dynamics (AIMD) simulations and displaced the atoms in the unit cell accordingly. This method can accurately reproduce the pre-edge intensities of measured Ti K-edge XANES of PbTiO$_3$ and SrTiO$_3$, but it worsens the agreement in the main edge and post edge between simulation and experiment due to the size limit of the unit cell. Zacharias, Patrick and Giustino~\cite{zacharias2015stochastic, zacharias2016one} proposed a ``one-shot" method that constructs a single, effective structure of the supercell, where atoms are displaced according to the magnitude of all the phonon modes weighted by the Bose-Einstein distribution. However, phonon calculations of the titanium compounds in the supercell of a few hundred atoms are quite expensive, plus some of the material systems in our study have significant anharmonicity that requires special treatment beyond the harmonic approximation.

In this study, we carry out the statistical average over snapshots of AIMD simulations to capture the thermal disorder effects. The contribution of isotropically averaged quadrupole transitions is examined. To account for many-body effects, we convolve the effective single particle (or static BSE) XANES spectrum with an \textit{ab initio} many-body spectral function, thereby including shake-up satellites. An energy-dependent Lorentzian broadening is further applied to simulated spectra to enable a quantitative comparison with measured spectra based on popular similarity metrics.

\section{Materials library}

Table~\ref{tab:9_materials} lists the nine Ti-containing compounds investigated in this study, which span a moderate range in space group, local site symmetry and the degree of local distortion. In all cases, the Ti atom is coordinated by six oxygen atoms, forming a TiO$_6$ octahedral motif. These octahedra exhibit varying degrees of distortion with lower site symmetries, except for SrTiO$_3$ which has an ideal, undistorted TiO$_6$ octahedron with the $O_h$ symmetry. To quantify the extent of structural distortion, Table~\ref{tab:9_materials} reports two metrics: the continuous symmetry measure (CSM: $\delta_c$)~\cite{pinsky1998continuous} and the displacement of the Ti atom from the geometric center of the octahedron, i.e. the center atom displacement ($\delta_d$). 

Besides SrTiO$_3$, two other materials (Rutile TiO$_2$ and CaTiO$_3$) are centrosymmetric, labelled by the inversion symmetry ($i$) in Table~\ref{tab:9_materials}. Rutile TiO$_2$ has a $D_{4h}$ local symmetry as a result of the axial elongation, with two long bonds (1.98~\AA) and four short bonds (1.95~\AA). It has a moderate distortion with a $\delta_c$ of 0.409. CaTiO$_3$ has a $C_i$ local symmetry. Three pairs of Ti-O bond lengths slightly deviate from the average value with a very small $\delta_c$ of 0.005. Anatase TiO$_2$ contains buckled TiO$_6$ octahedra with a $D_{2d}$ local symmetry. The pair of long Ti-O bonds (1.97~\AA) form a straight angle, while the two pairs of buckled short Ti-O bonds (1.94~\AA) has a O-Ti-O angle of 155.4 degrees. Although the Ti atom remains in the geometrical center of the octahedron, anatase TiO$_2$ has the largest $\delta_c$ of 3.031. Brookite TiO$_2$ forms a mixture of edge- and corner-sharing octahedra with the lowest local $C_1$ symmetry. Although the value of $\delta_d$ (0.18~\AA) is modest, its $\delta_c$ (1.238) is the second largest. Both BaTiO$_3$ ($C_{4v}$) and Ti$_2$O$_3$ ($C_3$) have moderate distortions with $\delta_c$ ranging from 0.1 to 0.3 and $\delta_d$ from 0.15~\AA{} to 0.16~\AA. Both FeTiO$_3$ and NiTiO$_3$ have $C_3$ local symmetry. The TiO$_6$ octahedron is edge sharing with (Ni/Fe)O$_6$ octahedra on one side of the three-fold axis and corner sharing with (Ni/Fe)O$_6$ octahedra on the other side. The Ti-O bond length on the edge-sharing side is 11\% to 12\% longer than that on the corner-sharing side. Because of that, FeTiO$_3$ and NiTiO$_3$ have a large $\delta_d$ of 0.29~\AA{} to 0.30~\AA. 

As for the chemical properties, Ti is in the +4 oxidation state for all compounds except Ti$_2$O$_3$, where it is +3. Due to its partially filled Ti 3$d$ bands, Ti$_2$O$_3$ is metallic in our DFT calculations.
Experimentally, Ti$_2$O$_3$ is nearly metallic, with only a very small band gap on the order of $0.1$ eV~\cite{li2018orthorhombic,chang2018c}. In the following discussion,  we focus the analysis on five representative compounds: SrTiO$_3$, BaTiO$_3$, rutile and anatase TiO$_2$, and NiTiO$_3$. Spectra for the remaining four compounds are in the Supporting Information.

\renewcommand{\arraystretch}{1.2} 

\begin{table}[ht]
\centering
\begin{tabular}{|c|c|cc|cc|c|}
\hline
\multirow{2}{*}{Material} & \multirow{2}{*}{Space group}  & \multicolumn{2}{c|}{Site symmetry} & \multicolumn{2}{c|}{Local distortion} & \multirow{2}{*}{Charge state} \\ 
\cline{3-6} 
 &   & \multicolumn{1}{c|} {Point group} & $i$ & \multicolumn{1}{c|}{$\delta_c$} & $\delta_d$ (\AA) & \\ \hline 
 
\textbf{SrTiO$_3$}~\cite{cif_SrTiO3}           & Pm$\bar{3}$m  & \multicolumn{1}{c|} {$O_{h} $} & T      
& \multicolumn{1}{c|} {0}      & 0  & 4+\\ \hline

CaTiO$_3$~\cite{cif_CaTiO3_liu1993x}  & Pbnm          & \multicolumn{1}{c|} {$ C_{i}$} & T     
& \multicolumn{1}{c|} {0.005}  & 0  & 4+\\ \hline

\textbf{TiO$_2$ (rutile)}~\cite{cif_TiO2}      & P4$_2$/mnm    & \multicolumn{1}{c|} {$D_{4h}$} & T     
& \multicolumn{1}{c|}  {0.409}  & 0 & 4+\\ \hline

\textbf{TiO$_2$ (anatase)}~\cite{cif_TiO2}     & I4$_1$/amd    & \multicolumn{1}{c|} {$D_{2d}$} & F    
& \multicolumn{1}{c|} {3.031}  &  0  & 4+\\ \hline

TiO$_2$ (brookite)~\cite{cif_TiO2}    & Pbca          & \multicolumn{1}{c|} {$C_{1} $} & F  
& \multicolumn{1}{c|}{1.238}  & 0.180  & 4+\\ \hline

\textbf{BaTiO$_3$}~\cite{cif_BaTiO3}           & P4mm          & \multicolumn{1}{c|} {$C_{4v}$} & F  
& \multicolumn{1}{c|} {0.102}  & 0.151 & 4+\\ \hline

Ti$_2$O$_3$~\cite{cif_Ti2O3}          & R$\bar{3}$c   & \multicolumn{1}{c|} {$C_{3} $} & F  
& \multicolumn{1}{c|}{0.275}  & 0.159 & 3+\\ \hline

\textbf{NiTiO$_3$}~\cite{cif_NiTiO3}    & R$\bar{3}$    & \multicolumn{1}{c|} {$C_{3} $} & F  
& \multicolumn{1}{c|} {0.883}  & 0.306 & 4+\\ \hline

FeTiO$_3$ ~\cite{cif_FeTiO3}          & R$\bar{3}$    & \multicolumn{1}{c|} {$C_{3} $} & F  
& \multicolumn{1}{c|} {0.914}  & 0.285  & 4+\\ \hline

\end{tabular}
\caption{Nine titanium-containing compounds used in the benchmark study. Five materials discussed in the main text are highlighted in bold font. Both crystal space group symmetry and local Ti site symmetry are listed. The column ``$i$" indicates whether the Ti site possesses inversion symmetry (T = true, F = false). The local distortion of the TiO$_6$ octahedron is quantified by the continuous symmetry measure (CSM: $\delta_c$) and the displacement of the Ti site from the geometric center ($\delta_d$).}
\label{tab:9_materials}
\end{table}

\section{Computational details}
\subsection{\textit{Ab initio} molecular dynamics simulations}
We performed AIMD simulations using density functional theory (DFT) implemented in the Vienna \textit{ab initio} Simulation Package (VASP)~\cite{VASP_PhysRevB.47.558, kresse1996efficient, kresse1996efficiency}. The exchange-correlation effects were treated with the Perdew-Burke-Ernzerhof (PBE) functional~\cite{GGA_PhysRevLett.77.3865} under the generalized-gradient-approximation. For NiTiO$_3$ and FeTiO$_3$, the simplified Hubbard $U$ correction was applied to $3d$ orbitals~\cite{PhysRevB.57.1505}, with effective $U$ values of 5.3 eV for Fe and 6.2 eV for Ni~\cite{HubbardU_PhysRevB.70.235121}. 
The statistics of atomic configurations were sampled with the canonical ensemble (NVT) at a temperature of 300 K. A Langevin thermostat was used with the friction coefficient set to 2 THz for all atoms. AIMD was performed in supercells with lattice vectors longer than 9~\AA. Depending on the material, the supercell contains 80 to 160 atoms. Lattice constants of the unit cell were set to their experimental values at 300 K and the supercell was fixed throughout the simulation. The AIMD simulation step was 1.25 fs. The first 2 ps of the trajectory were discarded to ensure equilibration. A 2.5 ps trajectory after the equilibration was used for the XANES simulation. 

\subsection{First-principles XANES simulations}
The x-ray absorption cross section can be calculated from Fermi's golden rule~\cite{de2008core}
\begin{equation} \label{eq:goldenrule}
  \mu(\omega)=4\pi^2 \alpha\, \omega\, \sum_f \left | M_{0,f} \right |^2  \delta(E_f-E_0-\omega),
\end{equation}
where $E_0$ and $E_f$ are the total energies of the many-body initial state $|\Psi_0 \rangle$ and final state $|\Psi_f\rangle$, $\alpha$ is the fine structure constant, and $\omega$ is the x-ray energy. Unless otherwise specified, we use atomic units throughout the rest of the paper. $M_{0,f}=\langle \Psi_f |\hat{O} | \Psi_0 \rangle $ is the transition matrix element with $\hat{O}$ the transition operator. Under the electric field of the photon beam, the dipole and quadrupole terms are given by $\hat{O}=\bs{e}\cdot \bs{r}+i/2 (\bs{e}\cdot \bs{r}) (\bs{q}\cdot \bs{r})$, where $\bs{e}$ and $\bs{q}$ are the polarization vector and
the wave vector of the photon beam, and $\bs{r}$ is the position of the electron.

The Ti K-edge XANES spectra were calculated by solving Eq.~\ref{eq:goldenrule} using the BSE method with the {\sc ocean} code \cite{OCEAN2022, OCEAN_PhysRevB.83.115106} and the excited-electron core-hole method with the VASP code~\cite{kresse1996efficient,kresse1996efficiency,karsai2018effects}. The finite-temperature effects in XANES are treated by performing a statistical average using AIMD snapshots. The same snapshots are used for {\sc ocean} and VASP calculations. Because the quadrupole term is not implemented in VASP, most of the results in the main text are taken from {\sc ocean} calculations. 

In the {\sc ocean} calculations, Kohn-Sham DFT orbitals were obtained using Quantum ESPRESSO \cite{QE_Giannozzi_2009,QE_Giannozzi_2017} with the PBE ONCV pseudopotentials~\cite{ONCV_PhysRevB.88.085117} from the PseudoDojo library \cite{PseudoDojo}.
To construct the BSE kernel, conduction bands up to approximately 70~eV above the conduction band minimum were included. The Brillouin zone was sampled on a regular k-point mesh with a linear density larger than 6 points per a.u.$^{-1}$ and the wavefunctions were down-sampled onto a real-space mesh with a linear density larger than 1 point per a.u.. The Ti core-hole potential was screened under the random-phase approximation (RPA). The dielectric response function was calculated using conduction bands spanning 120~eV and a k-point spacing of no less than 2.55 points per a.u.$^{-1}$. {\sc ocean} uses a combination of RPA and model dielectric screening \cite{OCEAN_screen}, and input values of the static dielectric constant $\epsilon_\infty$ were taken from the Materials Project \cite{Jain2013} for the insulators and set to 10000 for metals. In the {\sc ocean} quadrupole calculations, the powder average of the quadrupole term is performed by averaging over specially chosen photon orientations ({$\bs{e, q}$})~\cite{brouder1990angular}. The choice of photon orientations is given in Appendix A. The Haydock (Lanczos) method~\cite{lanczos1950iteration} is used to generate spectra with the contribution of each Ti atom and given photon orientation calculated separately. Convergence with respect to the number of iterations was checked according to integrated difference between the two spectra,
\begin{equation}
    \delta_{N,\Delta} = \frac{\int d\omega | \mu^{(N)}(\omega) - \mu^{(N-\Delta)}(\omega)|}{\sfrac{1}{2} \int d\omega \ \left (| \mu^{(N)}(\omega)| + |\mu^{(N-\Delta)}(\omega)| \right )} ,
\end{equation}
where $\mu^{(N)}$ is the spectrum generated using $N$ iterations. Iterations were continued until the spectrum at iteration $N$ and $N-5$ differed by less than $\delta=0.0001$. This convergence criteria required between 100 and 200 iterations for the systems and settings used here, with a constant Lorentzian broadening of 0.89~eV full width at half maximum (FWHM)~\cite{krause1979natural}. The core-hole binding energy is not calculated within {\sc ocean}, but relative core-level shifts between sites within each material are accounted for within a static core approximation \cite{OCEAN_screen,MengFanchen_PhysRevMaterials}.

The input files of VASP spectral simulations were generated using Lightshow~\cite{Lightshow_Carbone2023,MengFanchen_PhysRevMaterials}. We used the GW-type pseudopotentials, in order to obtain a good description of high-energy scattering states~\cite{MengFanchen_PhysRevMaterials}. A full core hole was included in spectral simulations and the core electron was placed at the bottom of the conduction band. The \emph{k}-point mesh used in the Brillouin zone sampling was determined using the effective crystal size method with the \code{cutoff} parameter set at {15~\AA} in Lightshow~\cite{MengFanchen_PhysRevMaterials}. The total number of bands ($n_b$) included in the spectral calculation depends on system size and the chosen energy range. Here we estimated $n_b$ based on the uniform electron gas model with the \code{e\_range} parameter set to 43~eV in Lightshow~\cite{MengFanchen_PhysRevMaterials}.
Note that unlike the BSE, the empty bands are eigenstates of the Kohn-Sham Hamiltonian in the presence of the core hole.

VASP spectral alignments across different absorbing sites within the same configuration, and across configurations, were performed by shifting the calculated edge according to
$E_{\mathrm{align}} = \left(E - E_{\text{CBM}}\right) + \left(E_{\mathrm{XCH}} - E_{\mathrm{GS}}\right)$~\cite{ENGLAND2011187,MengFanchen_PhysRevMaterials}, 
where $E_{\mathrm{XCH}}$ and $E_{\mathrm{GS}}$ are the total energies  of the excited state (core-hole system) and ground state, respectively. $E_{\text{CBM}}$ is the conduction band minimum of the excited state.
To match the experimental spectrum, the simulated spectrum was shifted in energy to achieve the highest cosine similarity score with the experimental spectrum.

We applied a Lorentzian broadening to simulated Ti K-edge spectra with an energy-dependent FWHM, $\Gamma(E_c) = \Gamma_{\text{core}} + \zeta \cdot (E_c - E_{\text{CBM}})$, where $\Gamma_{\text{core}}=0.89$ eV is the lifetime broadening of the Ti 1\textit{s} core hole~\cite{krause1979natural}. The linear term is a simple approximation of the quasiparticle life time broadening of an empty state with energy $E_c$.  $\zeta$ is an empirical parameter. For each simulation spectrum, we optimized $\zeta$ to get the best fit between simulation and experiment. 

\subsection{Many-body shake-up effects}
\label{shakeup}
Many-body shake-up effects were included by convolving the {\sc ocean} and VASP XANES spectra $\mu_{0}(\omega)$ with a core-hole spectral function~\cite{Calandra,JW1,JW2,Kas_PHC},
\begin{equation}
    \mu(\omega) = \int d\omega' A(\omega')\mu_{0}(\omega+\omega').
\end{equation}
The spectral function $A(\omega)=-1/\pi\ {\rm Im} [G(\omega)]$ was calculated within the cumulant approximation for the core-hole Green's function, which consists of both the quasiparticle peak and the shake-up satellites. In real time, $G(t)=G_0(t) \exp(C(t))$, where $G_0(t) = i\exp(-i\epsilon_c t)$ is the single-particle core-hole Green's function, and $C$ is the cumulant function, which incorporates the many-body shake-up effects \cite{Aryasetiawan, Guzzo, Kas_RC}. Note that the effect of  $G_0(t)$ causes only a constant shift of the spectrum $\mu(\omega)$ relative to $\mu_0(\omega)$. Since the calculation of core-level energies is already taken into account in  $\mu_0(\omega)$, we can safely set $\epsilon_c =0$ when calculating $A(\omega)$.  In this work, we calculate the cumulant within the real-time time-dependent density-functional theory approach \cite{Kas_RTC}, where the cumulant is obtained from the density $\delta\rho(t)$ induced by the sudden appearance of the core hole, i.e., 
\begin{align}
C(t) &= \int d\omega \frac{\beta(\omega)}{\omega^2}[e^{i\omega t}-i\omega t - 1],  \\
\frac{\beta(\omega)}{\omega} & ={\rm Re}\left[\int dt \ e^{-i\omega t} \int d^3 r V_c(\mathbf{r})\delta\rho(\mathbf{r},t)\right],
\end{align}
where $V_c(r)$ is the unscreened Coulomb potential of the core-hole. The function $\beta(\omega)$ characterizes the excitation spectrum of boson-like neutral valence excitations in the system, and is related to the loss function $\epsilon^{-1}(\omega)$. When a finite gap exists in $\beta(\omega)$, one can also relate the spectral function $A(\omega)$ to $\beta(\omega)$ in a more direct way by expanding in powers of $C(t)$, 
\begin{equation}
    A(\omega) = A_{qp}(\omega)\ast [1+A_1(\omega) + A_1(\omega)\ast A_1(\omega) +\cdots ],
\end{equation}
where $A_1(\omega)=\beta(\omega)/\omega^2$ is the single boson satellite spectral function, and $\ast$ denotes convolution in the above. Thus the first term in brackets denotes single boson excitation, the second double boson excitation, and so on. The weight of the quasi-particle  peak $A_{qp}(\omega)$ is thus reduced by the weight of all possible satellites.  
In general the spectral function has a sharp quasi-particle peak, as well as satellite structure corresponding to various many-body excitations such as charge-transfer or plasmons. As the spectral function is normalized, the effect of the convolution is to transfer weight from the quasiparticle spectrum $\mu_0$ to higher energies. The many-body satellites tend to be broad, and the effect on the XAS is essentially to reduce the size of the fine structure relative to the broad background function, similar to the amplitude reduction factor found in EXAFS analysis.

\section{Experimental XAS measurements}

Experimental spectra were collected at the National Institute for Standards and Technology's Beamline for Materials Measurement (BMM), beamline 6-BM at the National Synchrotron Light Source II. BMM uses a three-pole wiggler source. Light is collimated by a paraboloid mirror coated with 5\,nm of Rh on 30\,nm of Pt~\cite{Marcus:ie5001}. The collimated light is monochromated by a Si(111) double crystal monochromator. For these measurements, a flat, bare-silicon mirror provides harmonic rejection in an energy range around the Ti K edge energy. Commercially procured samples of each compound in our library of Ti standards were thoroughly mixed with polyethylene glycol powder and pressed into thin, 13\,mm diameter pellets. XAS data were collected at room temperature in transmission and in an energy range from 150\, eV below the Ti K edge energy ($\approx4966$\, eV) to $\approx$\,700\,eV above. These XANES data were normalized for comparison with theory using the \textsc{athena} program~\cite{RavelN05}. The measured data are archived at Ref.\ \onlinecite{mds2-4032}.

\section{Results and Discussion}

\subsection{Assignment of pre-edge features}


The subtle pre-edge features of 3$d$ transition metals K-edge XANES are closely related to the local chemistry of the absorbing site, such as symmetry, charge state, spin configuration and local crystal field. This subject has been well studied in the literature~\cite{cabaret1997determination, shirley2004ti, uozumi2004theory, de2007novel,cabaret2010first}. To set the stage for our later discussion, here we briefly summarize the main conclusions from the literature. Following Ref.~\cite{de2007novel}, the pre-edge consists of three types of excitations.
\begin{itemize}
    \item Local dipole transitions ($E_1$): $1s \rightarrow$ hybridized ($p$, $d$), where the absorbing atom's $4p$ state hybridizes with its 3$d$ state. This transition is forbidden for a centrosymmetric site, and only becomes allowed if the inversion symmetry is broken, either by static distortion or thermal fluctuation. 
    \item Non-local  dipole transitions ($E^{nl}_1$): 1\textit{s} $\rightarrow$ hybridized ($p$, $p_l$, $d_{n}$), where the absorbing atom's $4p$ state hybridizes with the 3$d$ states of neighboring transition metal atoms ($d_{n}$) via the ligand $p$ bands ($p_l$). 
    \item Local quadrupole transitions ($E_2$): 1\textit{s} $\rightarrow$ 3\textit{d} on the absorbing atom. The quadrupole peak is usually a few percent of the main peak~\cite{brouder1990angular}. Under the local octahedral geometry, the crystal field splits $3d$ orbitals into three $t_{2g}$ orbitals ($d_{xy}$, $d_{xz}$ and $d_{yz}$) at lower energy and two $e_g$ orbitals ($d_{x^2-y^2}$ and $d_{z^2}$) at higher energy. Therefore, $E_2$ transitions can give rise to two pre-edge peaks corresponding to $t_{2g}$ and $e_g$ final states. 
\end{itemize}
Because of the core-hole final-state effects, the local 3d states are pulled down in energy, making local transitions ($E_1$  and $E_2$) a few eV lower than the non-local dipole transitions ($E_1^{nl}$) that couple to the 3d band of non-excited Ti atoms~\cite{de2007novel,cabaret2010first}. This means that the first pre-edge peak (A$_1$) is comprised primarily of local transitions.


\subsection{Convergence on the number of snapshots used in thermal average}
\begin{figure}[htbp]
    \centering
    \includegraphics[width=2.5 in]{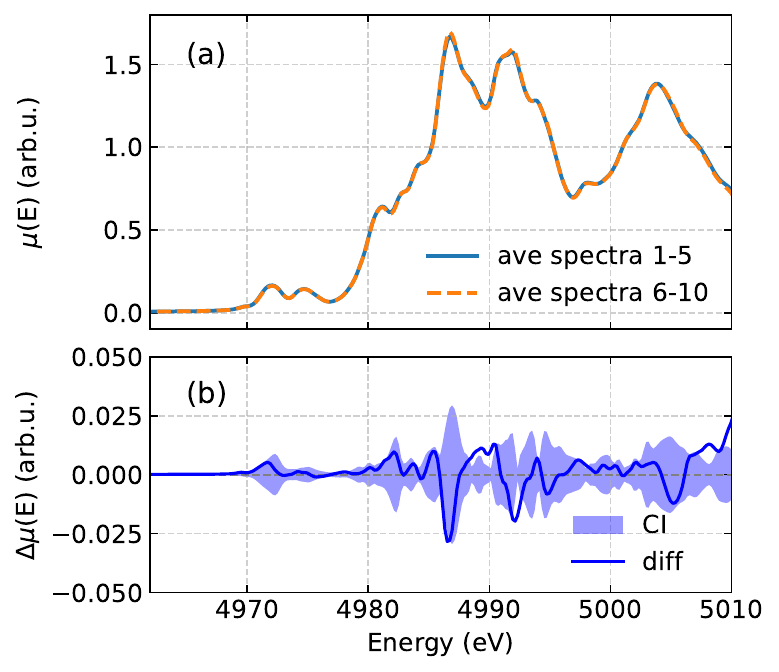}
    \caption{
    (a) Two thermal average VASP spectra calculated using two sets of 5 AIMD snapshots. The curves lie on top of each other. (b) The difference between the two averages plotted over the shaded confidence interval (CI). 
    }
    \label{fig:conv}
\end{figure}

We check the convergence of the statistical average of the thermal disorder effects. Default calculations are performed with 5 snapshots from AIMD simulations, with 0.5~ps between each snapshot. 
In each snapshot, the spectra are averaged over all Ti sites. We further compute the mean ($\bar{\mu}(\omega)$) of the average spectra ($\mu_i(\omega)$) from 5 snapshots. To quantify the uncertainty due to choosing only 5 snapshots, we define the confidence interval (CI) based on the variance  
\begin{align}
&s^2(\omega) = (N-1)^{-1}\sum_i^N \left( \mu_i(\omega) -\bar{\mu}(\omega) \right)^2,    \\
&\mu(\omega) \in \bar{\mu}(\omega) \pm \frac{C}{\sqrt{N}} s(\omega),
\end{align}
where $N$ is the number of the snapshots. 
We chose $C=2$, corresponding to an approximate 88\% confidence interval for the true spectrum $\mu$. 
Fig.~\ref{fig:conv} shows two mean VASP spectra calculated based on two different sets of 5 AIMD snapshots. We also compare the difference between these two averages with our calculated CI, and find that our calculated CI adequately captures the differences.
Fig.~\ref{fig:SIconv}a shows the mean {\sc ocean} spectrum and the CI using 5 snapshots. 
The variation of the average spectra of each snapshot is nearly invisible, confirming a good convergence. The same trend is found for VASP spectra as well as shown in Fig.~\ref{fig:SIconv}b. 

\subsection{Quadrupole contributions}
\label{quadrupole}

\begin{figure}[bhtp]
        \centering
        \includegraphics[width=6.5 in]{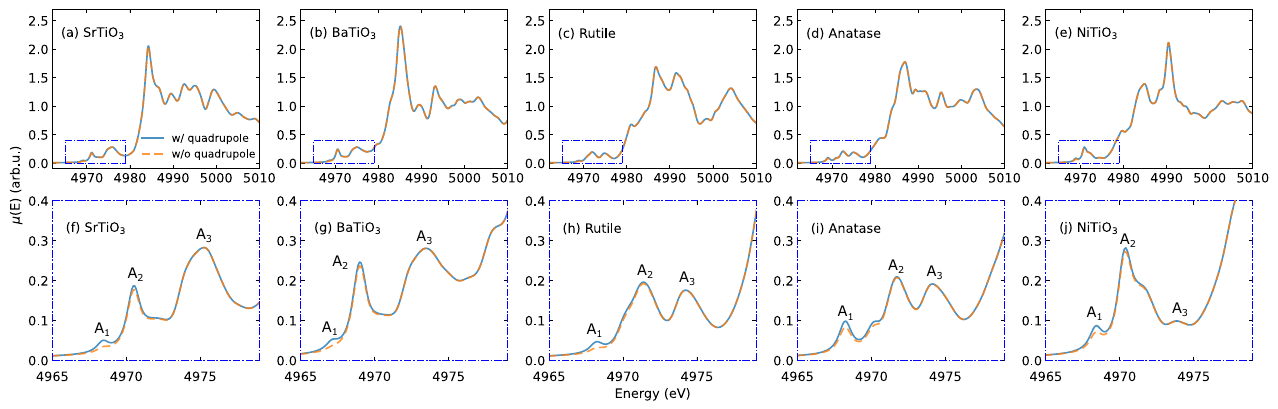}
        \vspace{-0mm}
        \caption{
        Quadrupole contribution to the pre-edge region of Ti K-edge XANES spectra. 
        (a--e) Full range XANES spectra. 
        (f--j) Zoom-in of the pre-edge region. 
        (a,f) SrTiO$_3$, (b,g) BaTiO$_3$, (c,h) Rutile TiO$_2$, (d,i) Anatase TiO$_2$, (e,j) NiTiO$_3$. 
        }
        \label{fig:quad_effect}
    \end{figure}

The effects of the quadrupole transitions ($E_2$) are shown in Fig.~\ref{fig:quad_effect} by comparing {\sc ocean} spectra with and without the quadrupole term. All the spectra are averaged over five AIMD snapshots. Since this is a comparison among simulations, only a constant Lorentzian broadening of 0.89 eV FWHM is applied. The many-body shake-up effects are not incorporated here, as we shall see later that it has negligible effects on pre-edge features. 

Quadrupole contributions do not affect the main edge and post edge,  which are dominated by $1s \rightarrow 4p$ dipole transitions. In the pre-edge region, peak A$_1$ mostly comes from $E_2$ transitions. For materials with site inversion symmetry (SrTiO$_3$, rutile TiO$_2$ and CaTiO$_3$; see also Fig.~\ref{fig:SI_quad_correction}). Peak A$_1$ is nearly invisible in the absence of $E_2$ transitions, and becomes clearly discernible when the quadrupole contribution is included. For materials with strongly buckled octahedra (anatase TiO$_2$) or a large $\delta_d$ value (NiTiO$_3$), peak A$_1$ already has significant intensity from $E_1$ transitions. In these cases, $E_2$ slightly increases the A$_1$ peak intensity. In addition, $E_2$ also slightly increases the intensity of the lower energy shoulder of the A$_2$ peak in anatase at 4970.2 eV. Overall, our results confirm that the quadrupole contribution is necessary to accurately reproduce the A$_1$ pre-edge peak intensities, particularly for centrosymmetric or nearly centrosymmetric absorbing sites.

\subsection{Thermal disorder effects} 
\label{thermal}

    \begin{figure}[htbp]
        \centering
        \includegraphics[width=6.5 in]{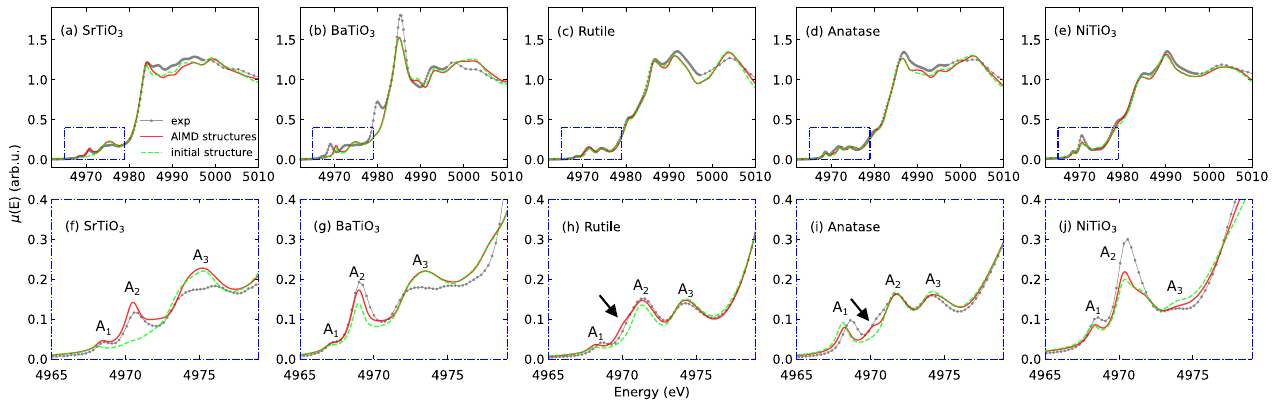}
        \vspace{-0mm}
        \caption{
        Thermal disorder effects on the Ti K-edge XANES spectra. 
        (a--e) Full range XANES spectra. 
        (f--j) Zoom-in of the pre-edge region; the simulated spectra are aligned to the experimental pre-edge region for better visualization. 
        (a,f) SrTiO$_3$, (b,g) BaTiO$_3$, (c,h) Rutile TiO$_2$, (d,i) Anatase TiO$_2$, (e,j) NiTiO$_3$. 
        Grey lines and markers: experimental spectra; red solid lines: simulation with thermal disorder; green dashed lines: simulation without thermal disorder. 
        Simulated spectra include quadrupole excitations, many-body shake-up effects, and an energy-dependent Lorentzian-broadening. 
        }
        \label{fig:thermal_effect}
    \end{figure}    

The thermal disorder effects are shown in Figs.~\ref{fig:thermal_effect} and \ref{fig:SI_thermal_correction} by comparing experimental spectra with simulated spectra with and without thermal fluctuations in the crystal structure. The thermal disorder effects are treated by averaging the XANES spectra from five AIMD snapshots at 300K, while spectra without thermal disorder are computed directly from the equilibrium experimental structure. All simulated spectra include quadrupole transitions and many-body shake-up effects, as well as an energy-dependent Lorentzian broadening to match with experimental spectra. 

Thermal disorder mostly affects the pre-edge features. Lattice vibrations can break the inversion symmetry in centrosymmetric systems, enabling otherwise forbidden local $E_1$ transitions and enhancing pre-edge peaks, especially on the A$_2$ peak. We notice that thermal disorder effects slightly enhance the A$_1$ peak in SrTiO$_3$, but have negligible impact on the A$_1$ peak in rutile TiO$_2$ and BaTiO$_3$. Interestingly, the A$_1$ peak intensity in anatase TiO$_2$ and NiTiO$_3$ slightly decreases after including thermal disorder because the averaging over AIMD snapshots damps sharp spectral features more than it enhances the $E_1$ transition. 

The A$_2$ peak in SrTiO$_3$ nearly entirely comes from the $E_1$ transition due to thermal disorder. This is consistent with the work of Cockayne \emph{et al.} based on AIMD and BSE calculations, where the A$_2$ peak of PbTiO$_3$ and SrTiO$_3$ is caused by the $t^{(1)}_{1u}$ phonon mode involving the asymmetric motion of the Ti center relative to its axial O neighbors~\cite{Cockayne_PhysRevB}. In rutile TiO$_2$ and BaTiO$_3$, the A$_2$ peak is attributed to combined $E_1$ and $E_1^{nl}$ transitions, while in the literature~\cite{cabaret2010first} the A$_2$ peak of rutile was assigned to $E^{nl}_1+E_2$ instead. The negligible $E_2$ contribution in the A$_2$ peak is evident from Fig.~\ref{fig:quad_effect}h. The discrepancy between our work and literature arises because the thermal disorder effects were not considered in Ref.~\cite{cabaret2010first}.
In BaTiO$_3$, thermal disorder significantly enhances the A$_2$ peak intensity, while in rutile TiO$_2$, thermal disorder manifests as a clear lower energy shoulder of the A$_2$ peak (arrow in Fig.~\ref{fig:thermal_effect}h). The strong buckling in anatase TiO$_2$ and the large $\delta_d$ value in NiTiO$_3$ give rise to a strong $E_1$ signature even without considering thermal fluctuations. Thermal disorder in these cases has a much weaker effect, including a lower energy shoulder of the A$_2$ peak in anatase (arrow in Fig.~\ref{fig:thermal_effect}i) and a slight increase of the A$_2$ peak intensity in NiTiO$_3$.



In comparison to the pre-edge intensities in the experiment, {\sc ocean} pre-edge spectra show excellent agreement in most of the cases with several visible discrepancies (Fig.~\ref{fig:thermal_effect}f-j). A more pronounced overestimation happens for the A$_3$ peak of SrTiO$_3$ ($\approx 4975$ eV) and BaTiO$_3$ ($\approx 4973$ eV). This is likely caused by BSE’s failure to account for charge-transfer processes that can result in effective screening of the Ti hole in the final state~\cite{Cockayne_PhysRevB}. However, BSE significantly underestimates the intensity of the A$_2$ peak in NiTiO$_3$. This could indicate a limitation in the underlying DFT description of the low-lying conduction band states, in particular, the unoccupied Ni 3{\it d} bands. The Hubbard U correction correctly favors a high-spin configuration for Ni$^{2+}$, but it is known that Hubbard corrections can have adverse effects on the unoccupied states probed by XAS \cite{Piccinin2019}.

In the main-edge region, thermal fluctuation results in a damping effect that broadens the spectra. After the energy-dependent Lorentzian broadening is applied, the effects of thermal disorder become less apparent in the main-edge region. Fig.~\ref{fig:SI_thermal_noBroaden} shows {\sc ocean} spectra with a 0.89 eV constant Lorentzian broadening instead of the energy-dependent Lorentzian broadening, where the thermally average spectra are smoother than the static spectra.

\subsection{Correlation of A1 peak with local distortion}
Thermal disorder plays an important role in pre-edge feature assignment, in particular for centrosymmetric systems, as thermal fluctuations break inversion symmetry and enable the otherwise forbidden $1s \rightarrow$ hybridized ($p$, $d$) dipole transitions. From the simulation side, in contrast to the quadrupole contribution that can be treated using the equilibrium structure, the effects of the thermal disorder requires performing molecular dynamics simulations and taking a statistical average over multiple snapshots. This significantly complicates the workflow of high-throughput XANES simulation and demands substantial computational resources. Therefore, an empirical description of the relationship between thermal disorder and local distortion not only provides physical insights but also facilitates practical data-driven XANES analysis. 

\begin{figure}[htbp]
        \centering
        \includegraphics[width=6.5 in]{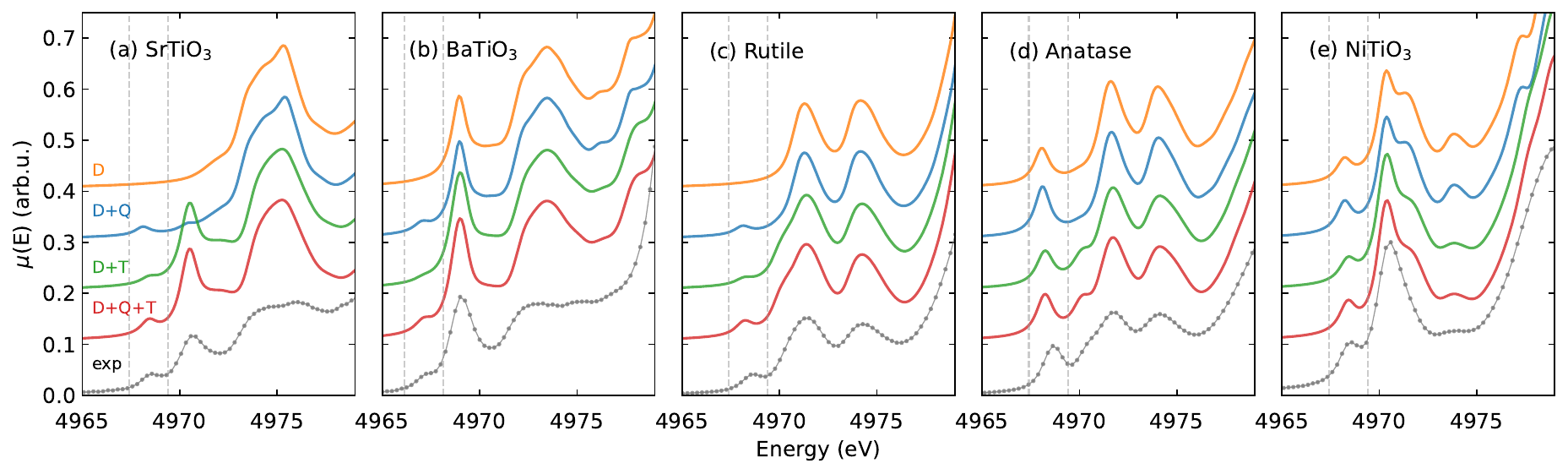}
        \caption{
        Pre-edge region showing the contribution of dipole transitions (D), quadrupole transitions (Q) and thermal disorder (T), offset vertically for clarity. 
        Simulations do not include the many-body shake-up effect and energy-dependent broadening in order to isolate pre-edge effects. 
        }
        \label{fig:pre-edge}
    \end{figure}

In this section, we quantify the effects of thermal disorder as a function of local distortions, characterized by $\delta_c$ and $\delta_d$. To exclude the contributions from the non-local excitations $E_1^{nl}$, we perform the analysis on the A$_1$ peak that primarily consists of local excitations ($E_1$ and $E_2$). In Figs.~\ref{fig:pre-edge} and \ref{fig:SI_pre_edge_noBroaden}, we plot the contributions including quadrupole transition (blue) and thermal disorder (green), along with the baseline curve of dipole transitions (orange) that neglects both. We measure the area under the A$_1$ peak using unbroadened spectra from dipole excitations without and with thermal disorder. The many-body shake-up effect and energy-dependent broadening are not included.

    \begin{figure}[hbtp]
        \centering
        \includegraphics[width=3.25 in]{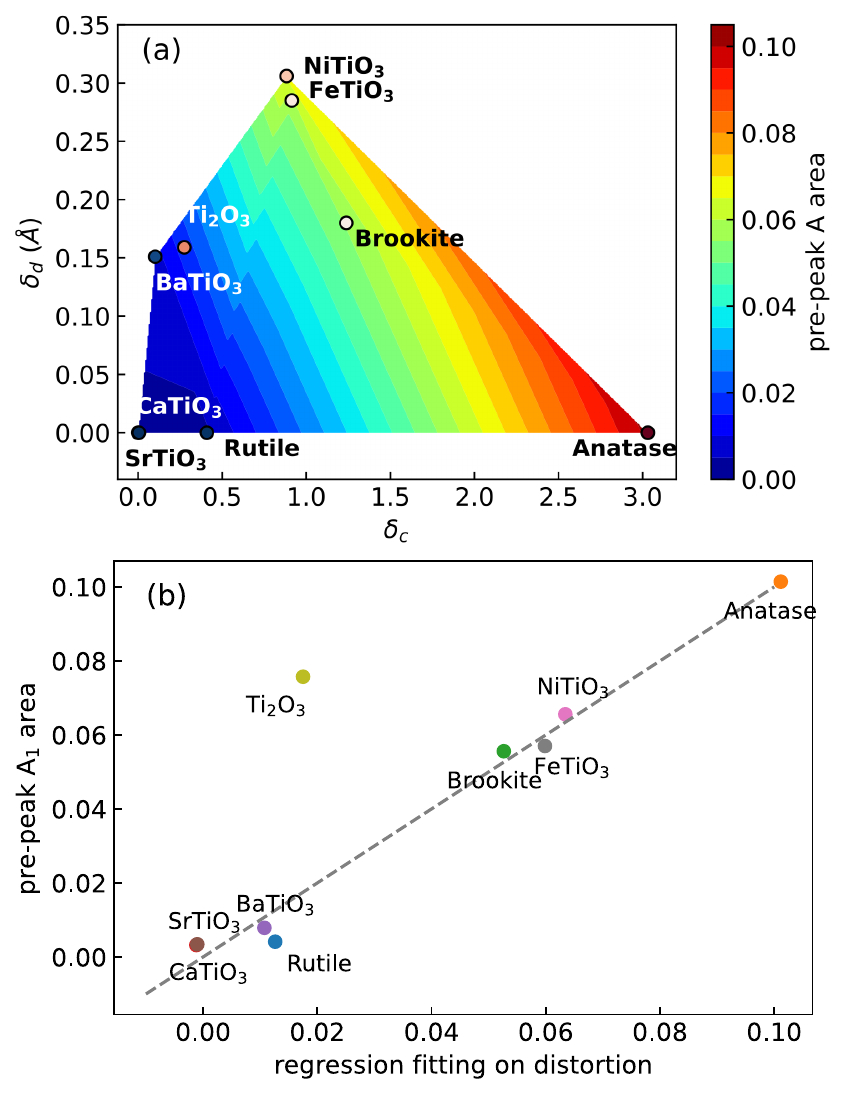}
        \caption{ (a) Contour plot of the peak A$_1$ area (dipole only; $A_{S}$) using the equilibrium structure with respect to static distortion characterized by $\delta_c$ and $\delta_d$. (b) Comparison of $A_{S}$ with the empirical relation based on the linear regression with $\delta_c$ and $\delta_d^2$ indicated by the dashed line.
        }
        \label{fig:sd}
    \end{figure}

We first look at the A$_1$ peak area according to the static distortion ($A_{S}$) measured from the dipole transitions using the equilibrium structure. In Fig.~\ref{fig:sd}a, we show the contour plot of $A_{S}$ with respect to $\delta_c$ and $\delta_d$, excluding Ti$_2$O$_3$. Overall, $A_{S}$ exhibits a smooth positive gradients along both $\delta_c$ and $\delta_d$.  The nine materials form five clusters in a wedge-shaped distribution. Along the perimeter, SrTiO$_3$, CaTiO$_3$ and rutile are at the lower left corner with small $\delta_c$ values ($\le 0.41$) and $\delta_d=0$. BaTiO$_3$ is located at the center-left, with $\delta_c=0.10$ and $\delta_d=0.15$. NiTiO$_3$ and FeTiO$_3$ are at the top center of the plot. They have the largest $\delta_c$ of 0.88 and 0.91, and moderate $\delta_d$ of 0.31 and 0.29. Anatase is located at the lower right with $\delta_d=0$, but the largest $\delta_c=3.03$. $A_{S}$ of anatase reaches a local maximum at $0.066$. Inside the wedge, brookite's local distortion is about half of the maximum, with $\delta_c=1.24$ and  $\delta_d=0.18$. $A_{S}$ of brookite is $0.057$, about 87~\% of anatase.
Ti$_2$O$_3$ appears as an outlier when overlaid on the contour plot, with a large $A_{S}=0.076$ that is 15~\% larger than anatase, but with relatively small $\delta_c=0.28$ and $\delta_d=0.16$. This discrepancy likely arises from several factors. First, Ti in Ti$_2$O$_3$ has a 3+ ($3d^1$) charge state, whereas Ti is formally 4+ ($3d^0$) in all other compounds. Furthermore, neighboring Ti atoms in Ti$_2$O$_3$ exhibit Ti-Ti dimerization, characterized by a single, short  Ti-Ti distance of 2.6~\AA. This strong next-nearest neighbor asymmetry is not reflected in the distortion metrics designed for individual Ti octahedra.

The relation between the pre-peak intensity and local distortion has been studied by others~\cite{cabaret2009origin,brouder2010effect,Cockayne_PhysRevB}. Here we generalize the Williams-Lax formalism~\cite{williams1951theoretical,lax1952franck} that treats the temperature-dependent optical absorption to XAS. For convenience, we switch the notation to the temperature-dependent imaginary part of the macroscopic dielectric constant $\epsilon_2(\omega; T)$, as $\mu(\omega;T) \propto \epsilon_2(\omega; T)$.
According to Zacharias, Patrick and Giustino~\cite{zacharias2015stochastic, zacharias2016one},
\begin{equation}
    \epsilon_2(\omega; T)=Z^{-1}\sum_n \exp(-E_n/k_BT)\langle \epsilon_2(\omega;x)\rangle_n,
\end{equation}
where $\langle \epsilon_2(\omega;x)\rangle_n$ is the expectation value evaluated at the nuclear quantum state $n$ at the atomic coordinates $x$ under the Born-Oppenheimer approximation. $E_n$ is the energy of the state $n$, $k_B$ is the Boltzmann constant, and $Z$ is the canonical partition function. Under the harmonic approximation, one can expand $\epsilon_2(\omega; T)$ in a Taylor series with respect to the phonon modes $x_\nu$~\cite{zacharias2016one},
\begin{equation}
    \epsilon_2(\omega; T)=\epsilon_2(\omega) + \frac{1}{2} \sum_\nu \frac{\partial^2 \epsilon_2(\omega; x)}{\partial x_\nu^2}\sigma_{\nu,T}^2 + \mathcal{O}(\sigma_{\nu,T}^4),
\end{equation}
where $\epsilon_2(\omega)$ is the XAS spectrum calculated at the equilibrium structure and $\sigma_{\nu,T}$ is the vibrational magnitude determined by the Bose-Einstein occupation of mode $\nu$.

In a centrosymmetric system, pre-peak A$_1$ of $\epsilon_2(\omega)$ vanishes, and $A_{S}$ is proportional to the second order of the small local distortion near the equilibrium,
\begin{equation}
    \epsilon_2(\omega; T)= \frac{1}{2} \sum_\nu \frac{\partial^2 \epsilon_2(\omega; x)}{\partial x_\nu^2}\sigma_{\nu,T}^2 + \mathcal{O}(\sigma_{\nu,T}^4). \label{eq:taylor}
\end{equation}
Eq.~\ref{eq:taylor} can also be derived for K-edge XAS pre-edge peaks using the dipole selection rule. Under the single-particle picture, we consider the initial state as the $1s$ orbital of the absorbing atom and the final state as the $p-d$ hybridization states,
\begin{equation}
    \epsilon_2(\omega;T)\propto |\langle 1s \ | \ \bs{e}\cdot \bs{r} \ | \ \textit{p-d}; T\rangle |^2.
\end{equation}
We expand the final state to the linear order: $| \textit{p-d}; T\rangle=|d\rangle + \delta |p; T\rangle$, where $\delta |p; T\rangle$ is a perturbation caused by inversion-symmetry breaking phonon modes. Because $\langle 1s \ | \ \bs{e}\cdot \bs{r} \ | \ d \rangle =0$, $\epsilon_2(\omega;T)\propto |\langle 1s \ | (\bs{e}\cdot \bs{r}) \ \delta |\ p; T\rangle |^2$. In general, we have $\epsilon_2(\omega; T)= \frac{1}{2} \sum_{\nu,\nu'} \frac{\partial^2 \epsilon_2(\omega; x)}{\partial x_\nu \partial x_{\nu'}}\sigma_{\nu,T} \sigma_{\nu',T} + \mathcal{O}(\sigma_{\nu,T}^4)$. In the limit of infinite number of normal modes, cross terms with $\nu \ne \nu'$ vanish~\cite{zacharias2016one} and Eq.~\ref{eq:taylor} is recovered.

Since the CSM ($\delta_c$) measures the similarity of a TiO$_6$ octahedron to a perfect octahedron based on the mean square of the bond length difference~\cite{pinsky1998continuous}, it corresponds to the second order in bond length distortion. Given $A_{S}$'s dependence on both $\delta_c$ and $\delta_d$ in Fig.~\ref{fig:sd}a, we perform a linear regression of $A_S$ on $\delta_c$ and $\delta_d^2$ excluding Ti$_2$O$_3$: $A_{S}=-0.001+0.034 \ \delta_c + 0.372 \ \delta^2_d$. The fitting yields a high coefficient of determination of $R^2=0.99$. Fig.~\ref{fig:sd}b shows the parity plot between the fitting and the target $A_{S}$. Clearly, the majority of the data points fall close to the exact solution, except for Ti$_2$O$_3$ showing a large deviation for reasons we previously discussed. 

    \begin{figure}[tbhp]
        \centering
        \includegraphics[width=3.25 in]{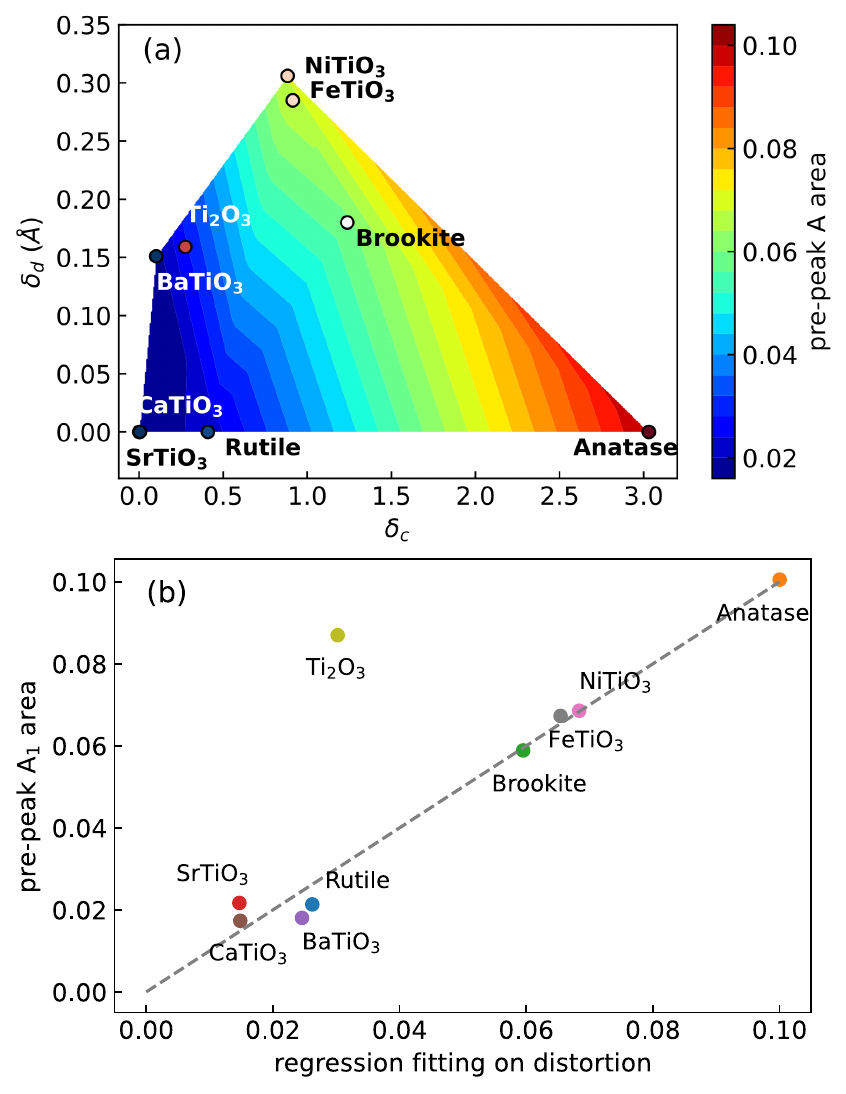}
        \caption{ (a) Contour plot of the peak A$_1$ area (dipole only; $A_{ST}$) with thermal average with respect to static distortion characterized by $\delta_c$ and $\delta_d$. (b) Comparison of $A_{ST}$ with the empirical relation based on the linear regression with $\delta_c$ and $\delta_d^2$ indicated by the dashed line.
        }
        \label{fig:sd+dd}
    \end{figure}

Next we quantify the effects of both static distortion and thermal disorder (ST). In Fig.~\ref{fig:sd+dd}a, we plot the contour plot of the area of pre-peak A$_1$ under dipole transition with thermal average ($A_{ST}$) with respect to $\delta_c$ and $\delta_d$. The pattern of $A_{ST}$ is essentially the same as $A_{S}$ in Fig.~\ref{fig:sd}a, but with overall larger values. Linear regression yields $A_{ST}=0.015+0.028 \ \delta_c + 0.308 \ \delta^2_d$ with $R^2=0.98$ as shown in Fig.~\ref{fig:sd+dd}b. Subtracting $A_{S}$ from $A_{ST}$, we obtain the contribution from thermal disorder, $A_{T}=0.016-0.006 \ \delta_c -0.064 \ \delta^2_d$. In systems with nearly perfect octahedral symmetry, thermal disorder breaks the inversion symmetry and results in a finite pre-peak A$_1$ area, corresponding to the positive constant term in $A_T$. In systems with finite structure distortions ($\delta_c$ and $\delta_d$), thermal disorder leads to a damping effect through the Debye-Waller factor, corresponding to the negative sign of the distortion terms. The net effect of $A_T$ depends on the competition of the opposite contributions from the constant term and distortion-dependent terms. At room temperature, we found that generally thermal disorder increases the area of pre-peak A$_1$, except for BaTiO$_3$, where the constant term and the distortion terms cancel each other.

\subsection{Many-body shake-up effects} 
\label{manybody}
\begin{figure}[htbp]
        \centering
        \includegraphics[width=6.5   in]{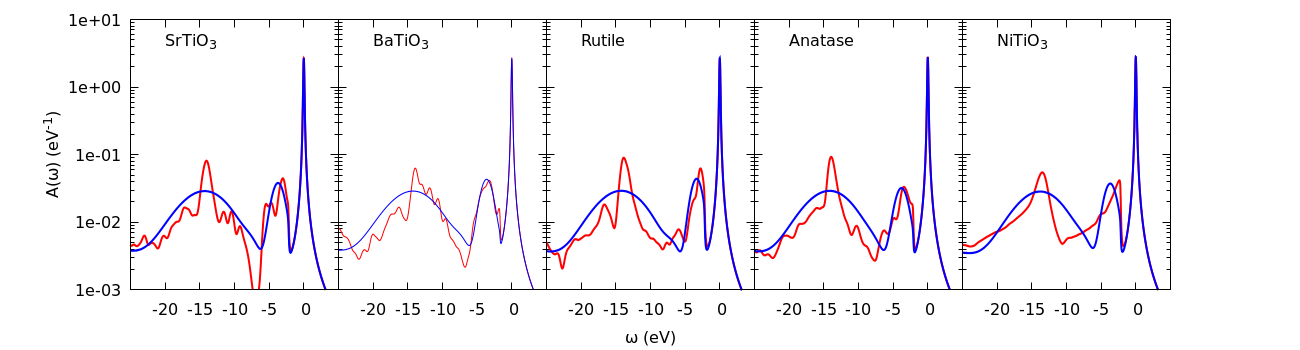}
        \vspace{-0mm}
        \caption{
        Model spectral function (blue) compared to the full spectral function (red) for each of the selected compounds. Note the similar behavior with a major satellite around $\omega\approx 13$ eV, and a low energy peak at $\omega\approx 5$ eV.
        }
        \label{fig:spfcns}
    \vspace{2mm}
    \end{figure}

The calculated core-hole spectral functions $A(\omega)$ are roughly similar for all of the materials studied here, as shown in Fig.~\ref{fig:spfcns} and Fig. S6, exhibiting a quasiparticle peak with differing degrees of low energy structure and a visible satellite structure at approximately 12~eV to 14~eV excitation energy, similar to what is seen in the XPS of SrTiO$_3$ and rutile TiO$_2$~\cite{JW1,JW2}. The high-energy satellites can be attributed to a zero crossing of the dielectric function, i.e., a plasmon. However, as noted previously \cite{kas_intrinsic}, these plasmons are quite different from those associated with free-electron metals, with the crossing originating from band-to-band transitions, and the excitations being much more localized. This is also consistent with the interpretation of these satellites as charge-transfer excitations.
In particular, the spherical hole potential causes monopole transitions between bonding and anti-bonding molecular orbitals of mixed ligand $p$ and metal $d$ character. The low-energy satellites are due to excitations between states comprised of O $p$ and metal $t_{2g}$, while the high-energy satellites originate from excitations between states of mixed O $p$ and metal $e_g$~\cite{JW2}.  As the effect on the XAS is only through a convolution with the spectral function, one can obtain a good approximation using a simple analytical model that assumes a two-peak form, i.e., 
\begin{equation}
\beta_{\rm model}(\omega)=\frac{a_1 \omega^2}{\Gamma_1\sqrt{\pi}}e^{-(\omega-\omega_1)^2/\Gamma_1^2}+\frac{a_2 \omega^2}{\Gamma_2\sqrt{\pi}}e^{-(\omega-\omega_2)^2/\Gamma_2^2}.
\end{equation}

The parameters of the model are extracted from the full calculation of $\beta(\omega)$ as follows. The excitation spectrum is split into low and high energy regions, with the division set at $1/2$ the energy of the maximum in $\beta(\omega)$ which is approximately 6~eV to 7~eV for the systems in question. For each region, the inverse first and inverse second moments of the model are matched to those of the full calculation of $\beta(\omega)$. In this way the quasiparticle renormalization $Z=\exp[-\int d\omega \beta(\omega)/\omega^2]$ (related to the EXAFS amplitude reduction factor $S_0^2$) is preserved. Finally, the widths were set to $1/3$ the peak position, e.g., $\Gamma_1 = 1/3\ \omega_1$. Table~\ref{tab:model} summarizes the extracted strengths ($a_1$ and $a_2$) and energies ($\omega_1$ and $\omega_2$) of the two-Gaussian model. The result confirms the similarity of the spectra, showing that both energy and intensity of the extracted Gaussian functions shows only small variation among the various systems, with the largest deviation being that of Ti$_2$O$_3$, which has a different Ti oxidation state ($3+$) than the rest of the systems ($4+$) and is metallic within our DFT calculations. 
Fig.~\ref{fig:spfcns} shows the comparison of this two-peak model to the full spectral functions. The difference between the convoluted XANES spectra with either the full spectral function or the two-peak model is negligible (see Fig.~\ref{fig:SI_xas_conv}).

\begin{table}[htb!]
\begin{center}
\begin{tabular}{|c || c| c| c| c|}
 \hline
 material & \hphantom{aa}$a_1$\hphantom{aa}
 &\hphantom{aa} $\omega_1$ \hphantom{aa}
 & \hphantom{aa}$a_2$ \hphantom{aa}
 & \hphantom{aa}$\omega_2$ \hphantom{aa}\\ [0.5ex]
\hline
\hline
SrTiO$_3$& 0.14 & 3.7 & 0.38 & 14.0 \\ 
 \hline
BaTiO$_3$ & 0.16 & 3.6 & 0.38 & 13.9 \\
 \hline
CaTiO$_3$ & 0.12 & 4.1 & 0.37 & 14.3 \\
\hline
rutile TiO$_2$& 0.15 & 3.4 & 0.37 & 13.9 \\
 \hline
anatase TiO$_2$ & 0.12 & 3.8 & 0.37 & 13.9 \\
 \hline
brookite TiO$_2$ & 0.12 & 3.8 & 0.38 & 13.8 \\ 
\hline
FeTiO$_3$ & 0.11 & 3.8 & 0.35 & 13.4 \\
\hline
NiTiO$_3$ & 0.13 & 3.7 & 0.35 & 13.5 \\  
\hline
Average$^*$ & 0.13 & 3.74 & 0.37 & 13.84 \\
\hline
Std. Dev.$^*$ & 0.02 & 0.20 & 0.01 & \hphantom{0}0.3 \\
\hline\hline
Ti$_2$O$_3$ & 0.91 & 1.4 & 0.4 & 11.67 \\ [0.5ex]
\hline
\end{tabular}
\caption{Parameters of the model excitation spectrum $\beta_{\rm model}(\omega)$ for each system. $^*$Average and standard deviation excluding Ti$_2$O$_3$.
\label{tab:model}}
\end{center}
\end{table}
\vspace{5mm}
 \vspace{5mm}
    \begin{figure}[htbp]
        \centering
        \includegraphics[width=6.5 in]{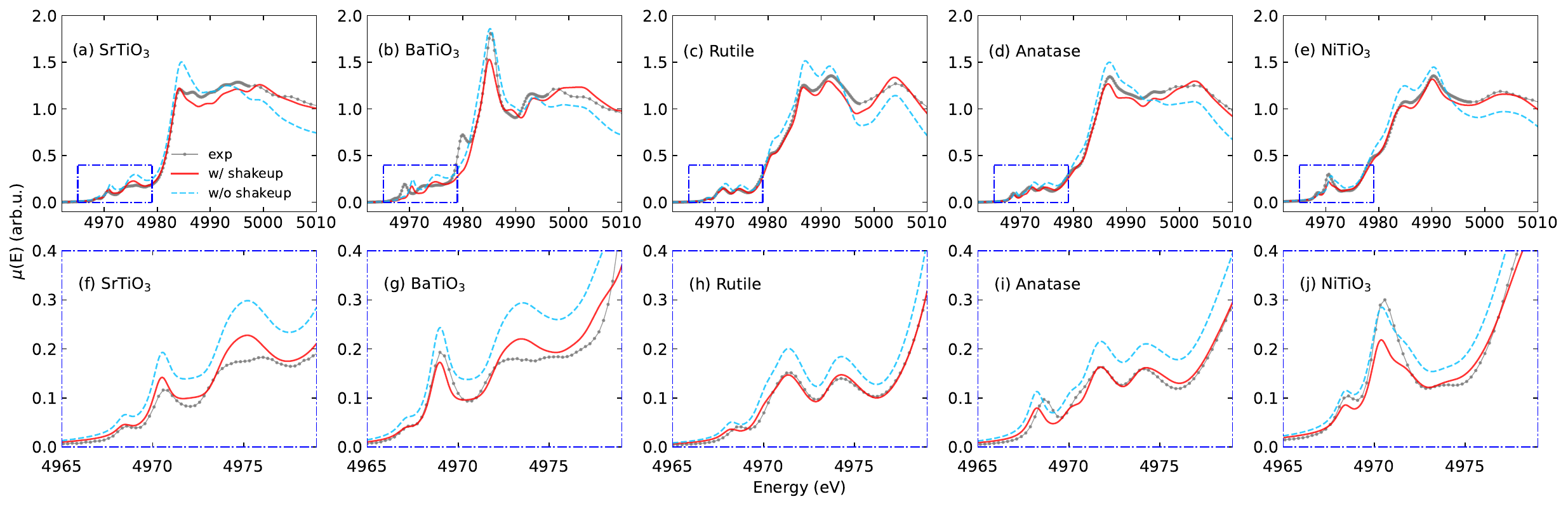}
        \vspace{-0mm}
        \caption{
        Effects of the many-body shake-up on Ti K-edge spectra. 
        (a--e) Full range XANES spectra. 
        (f--j) Magnified view of pre-edge region; the simulated spectra are aligned to the experimental pre-edge region to better visualization. 
        (a,f) SrTiO$_3$, (b,g) BaTiO$_3$, (c,h) Rutile TiO$_2$, (d,i) Anatase TiO$_2$, (e,j) NiTiO$_3$. 
        In each subplot, grey lines and markers represent experimental spectra, red solid lines represent simulated spectra with the shake-up effect, and blue dashed lines represent simulated spectra without the shake-up effect. 
        All simulations include the quadrupole contribution and thermal disorder effects. Simulated spectra are broadened by an energy-dependent Lorentzian function. 
        }
        \label{fig:shakeup_effect}
    \vspace{2mm}
    \end{figure}    

Figs.~\ref{fig:shakeup_effect} and \ref{fig:SI_shakeup_correction} compares experimental Ti K-edge XANES spectra with simulated spectra calculated with and without including the many-body shake-up effect. 
All simulated spectra include quadrupole transitions and thermal disorder effects and are broadened with an energy-dependent Lorentzian function to match experimental spectra. 
The many-body shake-up primarily affects the main-edge and post-edge region, while in the pre-edge region it only scales the intensity without changing the spectral shape.  
Overall the many-body shake-up effect redistributes the oscillator strength towards higher energy. Because the satellite peak $\omega_2$ in $A(\omega)$ is about $14$ eV below the quasiparticle peak, the main peak at around 4985 eV is weakened, as the $\omega_2$ peak is convoluted with the weak pre-edge below 4975 eV. In contrast, spectral features above $\approx$ 4995 eV are enhanced, because of the coupling between the peak at $\omega_2$ and the main peak. 

The effect on the spectral shape is particularly pronounced in rutile TiO$_2$ and NiTiO$_3$, where the main edge is a doublet with two peaks separated by about 4 eV (4987~eV and 4991 eV in rutile TiO$_2$; 4985~eV and 4990 eV in NiTiO$_3$). This doublet is in resonance with the $\omega_1$ satellite in $A(\omega)$ at about 4 eV, which transfers the oscillator strength from the low-energy peak to the high-energy peak of the doublet. In addition, the $\omega_2$ satellite causes the same effect. While the lower-energy peak of the doublet couples to the weak pre-edge through the $\omega_2$ peak, the higher energy peak couples to a much stronger features at $\approx$ 4977 eV in both systems.
Consequently, the shake-up effect significantly modifies relative peak intensity of the doublet, compressing the lower energy peak and enhancing the higher energy peak. Without the many-body shake-up effect, the low-energy peak is stronger than the high-energy peak in rutile TiO$_2$, and the trend is reversed when the many-body shake-up effect is included. As we can see in Fig.~\ref{fig:shakeup_effect}, including the shake-up effect in simulation generally yields a significant improvement in the spectral shape when compared with experiment.
 
We note that the core-hole final-state effect and the many-body shake-up effect have opposite impact on the spectral shape. In the core-hole final state, the electron--core-hole attraction pulls the oscillator strength to lower energy excitations, whereas the shake-up effect shifts the oscillator strength to higher energy. This leads to a partial cancellation.
The interplay of these competing effects can be clearly seen in rutile TiO$_2$. Under the initial-state rule, the main edge doublet has the same shape as the experiment with a stronger higher energy peak~\cite{poumellec1991electronic,MengFanchen_PhysRevMaterials}. Once the core-hole final-state effects are included at the BSE level, the lower energy peak becomes stronger, in contrast to experiment. Only when we include both core-hole final-state and many-body shake-up effects, the main peak restores the correct shape with a stronger higher energy peak as shown in Fig.~\ref{fig:shakeup_effect}c. Our analysis suggests that it is crucial to include all the key physics in the XANES simulations and neglecting one key effect could cause misinterpretation.

\subsection{Similarity quantification between simulation and experiment} 
In this section, we carry out a quantitative measure of the similarity between simulated and measured spectra. The goal is to systematically assess the accuracy of our computational methods in modeling Ti K-edge XANES for this small, representative material dataset. The results provide important references for first-principles XANES modeling. In the context of data-driven XAS analysis, our study helps illuminate the performance of ML models that are trained on simulation data and applied directly to experimental data. For completeness, we present the results for both {\sc ocean} and VASP (see VASP spectra Fig.~\ref{fig:SI_exp_vasp}). The {\sc ocean} spectra include the quadrupole contribution, thermal disorder effects, and many-body shake-up effects. The VASP spectra do not have the quadrupole contribution, as it is not implemented in the current version of VASP. Both results include the energy-dependent Lorentzian broadening. Three popular similarity metrics are considered, cosine similarity, Pearson correlation coefficient and Spearman correlation coefficient. 

\begin{table}[htb!]
\centering
\begin{tabular}{@{} c |c|c|c @{}} 
\toprule
material & Cosine & Pearson & Spearman \\
& {\sc ocean} \quad VASP & {\sc ocean} \quad VASP & {\sc ocean} \quad VASP \\
\midrule
SrTiO$_3$ & 0.999 \quad 0.999 & 0.997 \quad 0.997 & 0.942 \quad 0.962 \\
BaTiO$_3$ & 0.993 \quad 0.994 & 0.978 \quad 0.982 & 0.941  \quad 0.965 \\
CaTiO$_3$ & 0.999 \quad 0.999 & 0.996 \quad 0.997 & 0.974 \quad 0.972 \\
rutile TiO$_2$ & 0.999 \quad 0.999 & 0.996 \quad 0.996 & 0.984 \quad 0.981 \\
anatase TiO$_2$ & 0.999 \quad 0.999 & 0.998 \quad 0.998 & 0.981 \quad 0.985 \\
brookite TiO$_2$ & 0.999 \quad 0.999 & 0.998 \quad 0.998 & 0.981 \quad 0.983 \\
FeTiO$_3$ & 1.000 \quad 0.999 & 0.999 \quad 0.997 & 0.993 \quad 0.988 \\
NiTiO$_3$ & 1.000 \quad  0.999 & 0.999 \quad 0.997 & 0.995 \quad 0.991 \\
Ti$_2$O$_3$ & 0.999 \quad 0.998 & 0.997 \quad 0.991 & 0.948 \quad 0.931 \\
\hline
mean & 0.998 \quad 0.998  & 0.995 \quad 0.995 & 0.971 \quad 0.973 \\
$\sigma$ & 0.002 \quad 0.002 &  0.006 \quad 0.005 & 0.020 \quad 0.018 \\
\bottomrule
\end{tabular}
\caption{Similarity metrics between simulation and experiment using cosine similarity, Pearson correlation coefficient and Spearman correlation coefficient.}
\label{tab:similarity}
\end{table}

As shown in Table~\ref{tab:similarity}, both methods yield very accurate results as compared to experiment in the energy range between 4965 eV and 5005 eV, with mean values of all three metrics above 0.97. Since VASP spectra do not include quadrupole contributions, the nearly identical similarity scores in {\sc ocean} and VASP suggest that, because the quadrupole contributions only affect the pre-edge region, it does not affect the similarity measure of the spectral shape for the wide 40 eV range. Among the three metrics, the cosine similarity and Pearson correlation coefficient show higher scores with the mean of 0.998 (cosine) and 0.995 (Pearson) for both {\sc ocean} and VASP. The standard deviation ($\sigma$) is very small, occurring at the third decimal place. The Spearman correlation coefficient is slightly lower, with the mean and standard deviation of {\sc ocean } (VASP) at 0.971 (0.973) and 0.020 (0.018), respectively. Our results confirm that both {\sc ocean} and VASP spectra are accurate for quantitative Ti K-edge XANES analysis.

\subsection{Further investigation of the shoulder peak in Barium Titanate}
Despite the overall excellent agreement between simulation and experiment in the nine representative Ti compounds as shown in Fig.~\ref{fig:shakeup_effect}, we notice that our simulation results miss the pronounced shoulder peak in BaTiO$_3$ at $\approx$ 4980 eV. This discrepancy is also reflected in the low similarity scores in BaTiO$_3$: Cosine similarity of 0.993 ({\sc ocean}) and 0.994 (VASP), Pearson similarity of 0.978 ({\sc ocean}) and 0.982 (VASP), and Spearman similarity of 0.941 ({\sc ocean}) and 0.965 (VASP) in Table~\ref{tab:similarity}. Except for the lowest value of 0.931 in Spearman from Ti$_2$O$_3$ VASP, these are the lowest values in each category. Below we examine the possible cause of this discrepancy.

BaTiO$_3$ is a prototypical ferroelectric perovskite, whose ferroelectric properties and phase transition mechanism has been extensively studied. At low temperature, BaTiO$_3$ is rhombohedral with the polarization along the $<$111$>$ direction. As temperature increases, BaTiO$_3$ undergoes three phase transitions: it first transforms into the orthorhombic phase at 183 K with the polarization along $<$011$>$, and then into the tetragonal phase at 278 K with the polarization along $<$001$>$, and finally into the paraelectric cubic phase at 393 K~\cite{merz1949electric}. 

The microscopic origins of the BaTiO$_3$ phase transitions remain a subject of debate. In the displacive model, the polarization is associated with the Ti atom displacement as the transition temperature is approached from above due to the softening of displacement phonon mode~\cite{cochran1960crystal}. In the order-disorder model~\cite{bersuker1966origin,comes1968chain,comes1970desordre}, the Ti atom can occupy one of the eight local energy minima along the $<$111$>$ direction of the cubic cell. In the rhombohedral phase, the Ti positions exhibit a long-range order with the displacements all along the $<$111$>$ direction. In the orthorhombic phase, the Ti atoms become uncorrelated in one Cartesian direction. In the tetragonal phase, the Ti atoms lose the long-range correlation in two directions retaining only the correlation along $<$001$>$. Finally, in the cubic phase, all eight $<$111$>$ local minima are randomly occupied. While the displacive model can qualitatively explain the macroscopic and thermodynamic properties of BaTiO$_3$, there is growing evidence of the important role of the order-disorder picture in the phase transition of BaTiO$_3$ ~\cite{quittet1973temperature,ravel1998local,zalar2003nmr,hlinka2008coexistence,pugachev2012broken,tsuda2012nanoscale,tsuda2015two,tsuda2016direct,shao2017nanoscale,zhang2025real,gigli2022thermodynamics}. 


The 4980 eV shoulder is characteristic of Ti K-edge XANES of BaTiO$_3$ at temperatures spanning all the ferroelectric phases and the paraelectric cubic phase~\cite{ravel1998local,phaktapha2017temperature}. It appears in both single crystal samples~\cite{phaktapha2017temperature,ravel1998local} and epitaxial films grown using pulsed laser deposition~\cite{kato2021dielectric}. Because of that, the shoulder can be attributed to intrinsic materials properties of BaTiO$_3$, but the electronic origin of the shoulder remains obscure. Under an AC electric field, the time variation of the shoulder peak intensity is synchronized with the pre-edge A$_2$ peak~\cite{kato2021dielectric}.  In Zr-doped BaTiO$_3$, both the shoulder and A$_2$ peak intensities decrease with decreasing Ti concentrations~\cite{levin2011local,bootchanont2013investigation}. The shoulder peak intensity decreases monotonically with decreasing Ba concentrations in Ca-doped BaTiO$_3$~\cite{kato2021dielectric}, indicating that the shoulder excitations are coupled with Ba states. These observations suggest that the shoulder feature involves both local  Ti 3$d$ - 4$p$ mixing and ligand-mediated non-local excitations of Ti and Ba states outside the nearest-neighbor shell.

In addition to its phase transitions, another relevant material aspect of BaTiO$_3$ is its defect chemistry. Native point defects of BaTiO$_3$ include Ba vacancy, Ti vacancy, O vacancy (V$_{\text{O}}$), O interstitial, anti-sites, as well as defect complexes. V$_{\text{O}}$ often occurs in oxide perovskites as charge compensations of acceptor-type impurity cations, which are always dissolved into these materials during crystal growth~\cite{donnerberg2000ab}. According to DFT calculations, V$_{\text{O}}^{2+}$ is stable in a wide chemical potential range, especially under the oxygen deficient conditions where it is the most stable native defect~\cite{baker2018mechanisms,kanagawa2024first}. Therefore, V$_{\text{O}}^{2+}$ defects could have a non-negligible impact on the Ti K-edge XANES spectrum.

To augment the local structure motifs from the AIMD simulation of BaTiO$_3$, we further consider three types of structure models, including 1) local bond length and bond angle variation, 2) meta-stable Ti sites in disordered structures, and 3) the V$_{\text{O}}^{2+}$ defect.

\emph{Local distortion --} To investigate the effects of local bond-length and angle distortions on the spectrum we used the real-space multiple-scattering approach within the FEFF10 code \cite{RevModPhys.72.621,feff10_1,feff10_2} to calculate the XANES of clusters with the local TiO$_6$ octahedron modified as follows: 1) apical oxygen bond lengths modified by $\pm$10\%, and 2) two equatorial O-Ti-O angles reduced from 89.7 to 75 degrees, and 61.7 degrees. As shown in Fig.~\ref{fig:SI-feff_distortion}, a peak appears in all spectra between $4978$ and $4980$ eV, in good agreement with the location of the 4980 eV shoulder seen in the experimental results, but with much smaller amplitude. The local distortions do not modify this peak appreciably, indicating they can not explain the prominent shoulder at 4980 eV seen in experiment.

\emph{Metastable Ti sites --} To study the effects of the metastable Ti disorder, we performed structural optimization of the unit cell following a displacement of the Ti along the $\langle 011\rangle$ direction and along the $\langle 111\rangle$ direction. 
We compare the {\sc ocean} spectra of the two metastable structures with the undistorted unit cell in Fig.~\ref{fig:SI-ocean-metastable}. While both distortions show a growth in the low-energy shoulder of the main edge, neither is sufficient to explain the discrepancy with experiment. 

\emph{Oxygen vacancy --} In the experiment tetragonal structure (ideal structure), the apical Ti-O bond lengths are 1.829~\AA{} and 2.206~\AA, and the equatorial Ti-O bond length is 2.000~\AA~\cite{cif_BaTiO3}. To study the effect of the V$_{\text{O}}^{2+}$ defect, we removed one O apical atom (along the $<$001$>$ polarization direction) in the 3$\times$3$\times$3 tetragonal supercell and optimized the defect structure using the PBE functional. The two adjacent Ti atoms (Ti$_{sp}$) form square pyramids with five neighboring O atoms in the relaxed structure. The remaining apical Ti-O bond is shortened to 1.809~\AA, and the equatorial Ti-O bond length is shortened to 1.957~\AA. The Ti atom is offset from the equatorial plane with an O-Ti-O angle of $161.50$ degree.


As shown in Fig.~\ref{fig:pdos-isosurface}a, the VASP spectrum of the Ti$_{sp}$ site (indicated by the arrow in Fig.~\ref{fig:pdos-isosurface}b) has a pronounced shoulder about 1 eV higher than the shoulder in the experiment. Projected density of states (PDOS) and the isosurface of partial charge density of Ti$_{sp}$ in Fig.~\ref{fig:pdos-isosurface}b show a mixed electronic character of the shoulder.
The non-centrosymmetric square pyramid structure leads to strong Ti$_{sp}$ 3$d$ ($e_g$) - 4$p$ mixing on the absorber Ti that enables the $E_1$ transition. In addition, there is a major contribution from the non-local $E_1^{nl}$ excitation to the ligand-mediated states of Ti ($4s$, $4p$), O ($2p$), Ba ($4f$ and $5d$) and a small amount of neighboring Ti ($3d$, $4s$ and $4p$), in line with the experimental observations that the shoulder is sensitive to Ba and Ti concentration~\cite{kato2021dielectric,levin2011local,bootchanont2013investigation}. Since the shoulder has a significant Ti $3d$ character, it is not surprising the shoulder position in simulation is about 1 eV higher than experiment, a similar trend seen for the A$_2$ pre-edge peak in Fig.~\ref{fig:shakeup_effect}b. In comparison, the $3d$ - $4p$ mixing is insignificant on the Ti absorber site in the ideal structure as shown in Fig.~\ref{fig:pdos-isosurface}c. Regarding the Ti-O hybridization, Ti $4s$ does not participate in this energy range in the ideal structure, in sharp contrast to Ti$_{sp}$, where a strong Ti $4s$ peak overlaps with Ti $3d$, $4p$ and O $2p$. The Ti$_{sp}$ 4$p$ PDOS exhibits a clear split between 4980~eV and 4983 eV in the Ti$_{sp}$ 4$p$ - O 2$p$ hybridization, giving rise to the distinct shoulder feature, while in the ideal structure, the Ti $4p$ PDOS is more uniformly distributed between 4981 and 4984 eV, which does not produce a pronounced shoulder.

\emph{Effects of Ba 4f --} Our XANES simulations employ the PBE functional with Hubbard \(U\) corrections applied to the Fe and Ni \(3d\) states. However, local and semilocal DFT functionals are known to underestimate band gaps and can yield inaccurate band alignments. Consequently, the PBE electronic structure may not fully describe the relevant excitations in BaTiO$_3$, where hybridization among Ti, O, and Ba states can be important. In particular, the energy and localization of the unoccupied Ba $4f$ states may be inadequately described within PBE. {\it GW} calculations show that the self-energy correction shifts the Ti $3d$ conduction band upward by 1.8 eV and the O $2p$ valence band downward by 0.2 eV~\cite{sanna2011barium} against the PW91 functional~\cite{PhysRevB.46.6671,PhysRevB.33.8800} in BaTiO$_3$. However, the self-energy correction for the high-lying Ba $4f$ states remains unclear. 

\begin{figure}[bth!]
    \centering
    \includegraphics[width=3.4 in]{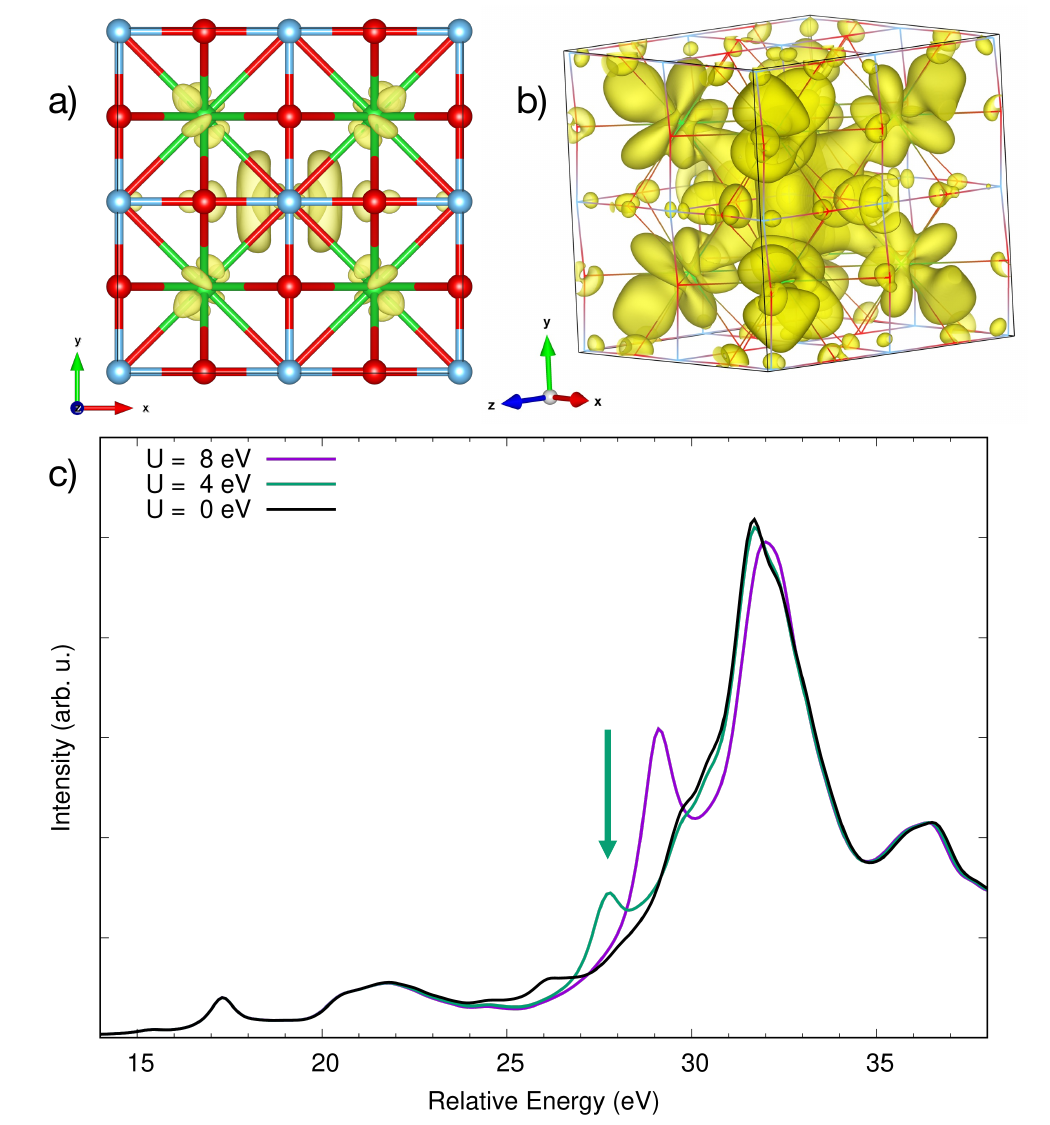}
    \caption{(a) The excited electron density for a dipole-limited x-ray excitation of the shoulder feature of the Ti K edge of BaTiO$_3$ (arrow in (c)) with the polarization along $\hat{x}$ and a Hubbard U=4~eV correction on the Ba 4{\it f} states (b) Same as (a), but at a smaller isosurface value. The wireframe style is used in the structure representation to give a clearer view of the electron density. (c) Comparison of calculated Ti K-edge XAS of BaTiO$_3$ for various Hubbard U corrections on the Ba 4{\it f}. Exciton density plots were made with VESTA \cite{VESTA}.}
    \label{fig:BTO+U}
\end{figure}

Given these significant self-energy corrections in the band alignment of BaTiO$_3$, a better understanding of the shoulder requires more accurate description of the band structure than PBE. While {\it GW} corrections are beyond the scope of this current work, we investigated the sensitivity of the spectra to Hubbard-U corrections on the Ba 4{\it f}. 
In the common case of applying a $+U$ correction to a partially filled manifold, e.g., 3{\it d} transition metal oxides, the correction serves to promote localization, dividing the manifold more cleanly between occupied and unoccupied states. Here, the Ba 4{\it f} states are all unoccupied, and the local +U serves to both increase their energy as well as promote their hybridization and delocalization. 
We show in Fig.~\ref{fig:BTO+U} that increasing the Hubbard $U$ correction primarily changes only the main discrepancy between the measured spectrum and our calculations. The feature increases in energy and intensity with larger values of U. While the increase in intensity improves agreement with experiment, the best agreement in terms of position with respect to the main edge is found near $U=0$~eV, indicating a limitation of using $+U$ alone to alleviate shortcomings in the exchange-correlation functional. At $U=8$ eV, a pronounced shoulder peak emerges with a peak intensity comparable to the experiment. However, the position of the shoulder is about 4 eV too high. A reasonable compromise is obtained with $U = 4$ eV, where a pronounced shoulder appears approximately 2 eV above the weak shoulder feature calculated without $U$.

This focused sensitivity indicates that the primary source of the discrepancy between our calculations and experiment is due to insufficient hybridization of the Ba 4{\it f} unoccupied states when using the semi-local PBE functional. The importance of the Ba 4{\it f}-like states on the shoulder feature is shown further by the isosurface plots of the photoelectron density of the shoulder peak at $U=4$ eV. In Fig.~\ref{fig:BTO+U}a, around the central absorbing Ti atom we see {\it p}-like density (for our choice of pseudopotential the 4{\it p} of Ti has a single node) as well as asymmetric density on the neighboring O {\it p} orbitals. Around the 8 nearest-neighbor Ba atoms we see distinct 6-lobed density features, indicative of significant role of the 4{\it f} orbitals to the shoulder peak. Ba $4f_{z^3}$ orbital plays a minor role in the shoulder, as its lobe points to the central Ti atom at a smaller isosurface value as shown in Fig.~\ref{fig:BTO+U}b.

In summary, our analysis suggests that an improved description of the 4980 eV shoulder feature in BaTiO$_3$ requires more accurate treatment of the correlation effects of Ba $4f$ states, using DFT+U or GW self-energy correction. Defects, such as Ti square pyramid motifs in oxygen deficit BaTiO$_3$, may also contribute to the shoulder. 

\section{Conclusion}

In this work, we benchmark \textit{ab initio} simulations of Ti K-edge XANES spectra against experimental measurements for nine Ti-containing compounds. 
In addition to the band theory treatment of the core-hole final state effects, three key physical effects are incorporated, including electric quadrupole transitions, thermal fluctuations, and many-body shake-up. We find that quadrupole transitions and thermal disorder predominantly affect the pre-edge region, whereas many-body shake-up has a strong influence on both main-edge and post-edge regions. 

While quadrupole transitions ($E_2$) dominate the A$_1$ peak in centrosymmetric systems (SrTiO$_3$, rutile TiO$_2$ and CaTiO$_3$), their contribution becomes insignificant in materials with strongly buckled octahedra (anatase TiO$_2$) or a large center displacement (NiTiO$_3$), where local dipole transitions ($E_1$) prevail. Thermal disorder effects are essential to obtain the accurate A$_2$ peak shape and intensity, especially in centrosymmetric systems (SrTiO$_3$, rutile TiO$_2$ and CaTiO$_3$) and systems with small local distortions (e.g., BaTiO$_3$). In systems with strong local distortions (e.g., buckled anatase TiO$_2$ and NiTiO$_3$ with a large center displacement), thermal disorder effects can give rise to a lower energy shoulder of the A$_2$ peak or a slight increase of the A$_2$ peak intensity. There are a few remaining discrepancies between simulation and experiment, including the overestimation of the A$_3$ peak intensity in SrTiO$_3$ and BaTiO$_3$, and the underestimation of the A$_2$ peak intensity in NiTiO$_3$. Fully resolving these discrepancies requires further investigations. The contribution of the thermal disorder effects to the A$_1$ peak area can be very well explained by local distortion descriptors, where the A$_1$ peak area fits accurately to a linear combination of the continuous symmetry measure and the square of the center displacement.  


The many-body shake-up effect strongly influences the spectral shape of both the main edge and post edge. This effect is accurately captured by the cumulant method, where the satellite features in the core-hole spectral function can be approximated by two satellite peaks. Overall, the shake-up effect transfers the oscillator strength to higher energy, which increases the post edge intensity relative to the main edge. It has the opposite effects of the BSE, which enhances low energy excitations. For this reason, the many-body shake-up effect needs to be considered together with core-hole final state effects to obtain the correct balance at the main edge and the post edge. The shake-up effect becomes more crucial, when the main edge has two peaks such as rutile TiO$_2$. The convolution with the core-hole satellite peaks can enhance the high energy peak against the low energy peak, thereby significantly modifying the relative peak intensity. In this work, we accurately reproduce the relative doublet peak intensity in the main edge of rutile TiO$_2$ by including the shake-up effects, resolving an important issue in the previous study~\cite{MengFanchen_PhysRevMaterials}.



The accuracy of the simulation is validated based on similarity scores between simulated and measured spectra of nine titanium compounds in an energy range of 40 eV. Overall, high similarity scores are achieved for both {\sc ocean} (Cosine similarity above 0.993, Pearson similarity above 0.978, and Spearman similarity above 0.941) and VASP spectra (Cosine similarity above 0.994, Pearson similarity above 0.982, and Spearman similarity above 0.931). The results confirm the predictive power of \emph{ab initio} XANES simulations for quantitative spectral analysis, especially regarding the spectral line shape. The remaining issue of the missing shoulder in BaTiO$_3$ does not have a simple origin and warrants further investigation of relevant structure models with higher level electronic structure theory.

Our work establishes the workflow to build accurate simulated Ti K-edge XANES databases. This workflow can be generalized to a wide range of materials, as far as their 1\emph{s} excitations are governed by weakly correlated physics that can be described by band theory. We envision that our XANES benchmark study will have a significant impact on XAS analysis using both traditional first-principles methods and an AI-assisted workflow, where high quality simulated spectral data are essential for training high-fidelity machine learning models.

\section*{Acknowledgments}

This work used the Theory and Computation resources at the Center for Functional Nanomaterials and beamline 6-BM of the National Synchrotron Light Source II, both of which are U.S. Department of Energy (DOE) Office of Science User Facilities, at Brookhaven National Laboratory under Contract No. DE-SC0012704. This research used resources of the National Energy Research Scientific Computing Center (NERSC), a Department of Energy User Facility using NERSC awards No. BES-ERCAP0028324, No. BES-ERCAP0032137, and No. BES-ERCAP0036930. 
JJK aknowledges support by the Theory Institute for Materials and Energy Spectroscopy at SLAC (Grant No. FWP100291), which is funded by U.S. Department of Energy Office of Science BSE DMSE Contract No. DE-AC02-76SF0051.
We thank Dr.\ Martin Stennett of the University of Sheffield for providing our library of Ti standard specimens. We thank Mark Hybertsen, David Prendergast and Fabrice Roncoroni for helpful discussions.

Certain software is identified in this paper in order to specify the experimental procedure adequately.  Such identification is not intended to imply recommendation or endorsement of any product or service by NIST, nor is it intended to imply that the software identified is necessarily the best available for the purpose.

\newpage
\appendix
\section{Isotropic average of quadrupole spectra}
\subsection{General formula}
The quadrupole-only contribution, $\sigma^Q$ can be expressed as complex-valued elements of a spherical tensor $\sigma^Q(l,m)$ for $l=0,2,4$ and $m=-l,...,l$. The contribution of each as a function of the Euler angles is given in the work of Brouder~\cite{brouder1990angular}, Eq.~A38, reproduced here as Eq.~\ref{eq-sigmaQ}. 
\begin{align}
\sigma^Q(\hat\epsilon,\hat k)\;=\;&\sigma^Q(0,0)
+\sqrt{\frac{5}{14}}\,(3\sin^2\theta\,\sin^2\psi-1)\,\sigma^Q(2,0) \nonumber \\
&+2\sqrt{\frac{15}{7}}\;\sin\theta\,\sin\psi\,
\Bigl[\,
\cos\theta\,\sin\psi\,\bigl(\sigma^{Q r}(2,1)\cos\phi
  +\sigma^{Q i}(2,1)\sin\phi\bigr) \nonumber \\[-2pt]
&\qquad\quad{}+\cos\psi\,\bigl(\sigma^{Q r}(2,1)\sin\phi
  -\sigma^{Q i}(2,1)\cos\phi\bigr)\Bigr] \nonumber \\[6pt]
&+\sqrt{\frac{15}{7}}\,
\Bigl[\,
(\cos^2\theta\,\sin^2\psi-\cos^2\psi)
\bigl(\sigma^{Q r}(2,2)\cos2\phi+\sigma^{Q i}(2,2)\sin2\phi\bigr) \nonumber \\[-2pt]
&\qquad\quad{}+2\,\cos\theta\,\sin\psi\,\cos\psi\,
\bigl(\sigma^{Q r}(2,2)\sin2\phi-\sigma^{Q i}(2,2)\cos2\phi\bigr)\Bigr] \nonumber \\[6pt]
&+\frac{1}{\sqrt{14}}\,
\bigl[\,35\sin^2\theta\,\cos^2\theta\,\cos^2\psi
+5\sin^2\theta\,\sin^2\psi-4\bigr]\,\sigma^Q(4,0) \nonumber \\[6pt]
&+\sqrt{\frac{10}{7}}\;\sin\theta\,
\Bigl[\,
(14\cos^2\theta\,\cos^2\psi+8\sin^2\psi-7)\,\cos\theta\,
\bigl(\sigma^{Q r}(4,1)\cos\phi+\sigma^{Q i}(4,1)\sin\phi\bigr) \nonumber \\[-2pt]
&\qquad\quad{}-(7\cos^2\theta-1)\,\sin\psi\,\cos\psi\,
\bigl(\sigma^{Q r}(4,1)\sin\phi-\sigma^{Q i}(4,1)\cos\phi\bigr)\Bigr] \nonumber \\[6pt]
&-2\sqrt{\frac{5}{7}}\,
\Bigl[\,
(7\sin^2\theta\,\cos^2\theta\,\cos^2\psi
 +\cos^2\theta\,\sin^2\psi-\cos^2\psi)\,
\bigl(\sigma^{Q r}(4,2)\cos2\phi+\sigma^{Q i}(4,2)\sin2\phi\bigr)\nonumber \\[-2pt]
&\qquad\quad{}-(7\sin^2\theta-2)\,\cos\theta\,\sin\psi\,\cos\psi\,
\bigl(\sigma^{Q r}(4,2)\sin2\phi-\sigma^{Q i}(4,2)\cos2\phi\bigr)\Bigr] \nonumber \\[6pt]
&+\sqrt{10}\,\sin\theta\,
\Bigl[\,
(1-2\cos^2\theta\,\cos^2\psi)\,\cos\theta\,
\bigl(\sigma^{Q r}(4,3)\cos3\phi+\sigma^{Q i}(4,3)\sin3\phi\bigr) \nonumber \\[-2pt]
&\qquad\quad{}+(3\cos^2\theta-1)\,\sin\psi\,\cos\psi\,
\bigl(\sigma^{Q r}(4,3)\sin3\phi-\sigma^{Q i}(4,3)\cos3\phi\bigr)\Bigr] \nonumber \\[6pt]
&+\sqrt{5}\,\sin^2\theta\,
\Bigl[\,
(\cos^2\theta\,\cos^2\psi-\sin^2\psi)\,
\bigl(\sigma^{Q r}(4,4)\cos4\phi+\sigma^{Q i}(4,4)\sin4\phi\bigr) \nonumber \\[-2pt]
&\qquad\quad{}-2\,\cos\theta\,\sin\psi\,\cos\psi\,
\bigl(\sigma^{Q r}(4,4)\sin4\phi-\sigma^{Q i}(4,4)\cos4\phi\bigr)\Bigr]
\label{eq-sigmaQ}
\end{align}
where the polarization and momentum unit vectors are both defined from the set of angles
\begin{align}
\hat{\varepsilon} &= (\sin\theta \cos \phi,\sin \theta \sin \phi,\cos \theta) \\
\hat{k} &= ( \cos\theta \cos\phi \cos\psi-\sin\phi \sin\psi,
\cos\theta \sin\phi \cos\psi + \cos\phi \sin\psi, 
-\sin\theta \cos\psi )
\end{align}

\if{0}
In the codes we have used here, spectra are generated by considering specific cartesian polarization and momentum vectors, often chosen to match experimental conditions. For a disordered or powder sample, the isotropic average is needed. In the case of dipole-limited spectra, averaging over the spectra generated from any three orthogonal polarizations is sufficient. For the quadrupole term, we carried out numerical optimization to design a set of polarizations and momenta that give the correct isotropic average for both dipole and quadrupole components
\begin{align}
&\sum_i^N \frac{\sigma^Q(\hat{\epsilon_i},\hat{k_i})}{N} \approx \sigma^Q(0,0) \\
&\sum_i^N \frac{(\epsilon_i\cdot \hat{x})^2}{N} \approx \sum_i^N \frac{(\epsilon_i\cdot \hat{y})^2}{N} \approx \sum_i^N \frac{(\epsilon_i\cdot \hat{z})^2}{N} \approx \frac{1}{3}
\end{align}
We determined a non-unique solution using 5 sets of directions that is numerically exact within the tolerances of the optimization, with errors less than $10^{-20}$. The optimization script is available in the SI, while the 5 directions are shown in Table~\ref{tab:5photons}.
\begin{table}[]
    \centering
    \begin{tabular}{c|c c c }
    $\hat{\epsilon}_1$ &  $-$0.91388674931755841667 & $-$0.38103254337885177951 &  $\phantom{-}$0.14009000788075350010 \\
    $\hat{k}_1$ &   $\phantom{-}$0.15906369646387180678 & $-$0.01859238773662542490 &  $\phantom{-}$0.98709323956022525085 \\
    \hline
$\hat{\epsilon}_2$ &  -0.32979387629775339157 &  0.48744973674909068801 &  0.80847309992339529969 \\
$\hat{k}_2$ &   0.56335004038734553731 & -0.58558098984498667739 &  0.58286502410741868239 \\
    \hline
$\hat{\epsilon}_3$ &  -0.40873684751330899373 & -0.31387292804666514660 &  0.85698189859780291340 \\
$\hat{k}_3$ &   0.16635472797265664420 & -0.94889333997811432653 & -0.26819309056409739770 \\
    \hline
$\hat{\epsilon}_4$ &  -0.72576069778373385995 & -0.62057361903794472075 & -0.29691714821918854457 \\
$\hat{k}_4$ &  -0.68790804214612095467 &  0.65003908258436385361 &  0.32284937147773614040 \\
    \hline
$\hat{\epsilon}_5$ &  -0.17005667961684027352 & -0.89456458128505909903 &  0.41332183057271755217 \\
$\hat{k}_5$ &   0.90725452281117130824 &  0.02159894847302956839 &  0.42002704230029434378 \\
    \end{tabular}
    \caption{An optimized solution for 5 sets of polarization and momentum that yield an isotropic x-ray spectra at both dipole and quadrupole order}
    \label{tab:5photons}
\end{table}
\newpage

\section{Isotropic average of quadrupole spectra}
\fi

Similarly, the dipole-only contribution $\sigma^D$ has elements $l=0,2$ (Eq. 4.7 in Ref.~\onlinecite{brouder1990angular}), reproduced here as Eq.~\ref{eq-sigmaD},
\begin{align}
\label{eq-sigmaD}
    \sigma^D(\hat{\epsilon}) &= \sigma^D(0,0) - \sqrt{3} \sin^2\theta \left[ \cos2\phi \sigma^{Dr}(2,2) + \sin2\phi \sigma^{Di}(2,2) \right] \nonumber \\
    &+ 2 \sqrt{3} \sin\theta\cos\theta\left[ \cos\phi \sigma^{Dr}(2,1) + \sin\phi \sigma^{Dr}(2,1) \right] \nonumber \\
    &- 1/\sqrt{2} \left(3 \cos^2\theta -1 \right) \sigma^D(2,0)
\end{align}

Our goal is to determine sets of Euler angles such that the average causes all the geometric prefactors for all $l\ne0$ terms to vanish for both $\sigma^D$ and $\sigma^Q$ regardless of the values of the spherical tensors.  
If we have a set of 3 Euler angles that is given by $(\theta_A, \phi_A+\frac{2n\pi}{3},\psi_A)$, for $n=0,1,2$, then the average gets rid of all $m\ne0,3$ terms. This leaves five undetermined terms, three $m=0$  $\{\sigma^D(2,0), \sigma^Q(2,0), \sigma^Q(4,0) \}$ and two $m=3$ $\{\sigma^{Qr}(4,3), \sigma^{Qi}(4,3) \}$. We therefore include a second set of 3 Euler angles $(\theta_B, \phi_B+\frac{2n\pi}{3},\psi_B)$, giving us a total of six variables with which to zero out five pre-factors. 
We can set $\phi_A=0$ and then find an analytical solution for canceling out the $m=3$ terms. 
The geometric prefactor for the real/imaginary $\sigma^Q(4,3) =  \mathrm{Re/Im} \Big[ C(\theta,\psi)  e^{3i\phi} \Big] $.
\begin{align}
&C(\theta,\psi)\equiv\sqrt{10} \sin\theta\ \bigl[ (1-2\cos^2\theta\,\cos^2\psi)\,\cos\theta\,
 +-(3\cos^2\theta-1)\,\sin\psi\,\cos\psi\,
\bigr]
\end{align}
With our sets of Euler angles we have the constraint that the average of $C$ must be zero
\begin{equation}
    \sfrac{1}{2}\, C(\theta_A,\psi_A) + \sfrac{1}{2}\, C(\theta_B,\psi_B) e^{i 3\phi_B} = 0 \quad .
\end{equation}
Therefore, with the constraint that $|C(\theta_A,\psi_A)|= |C(\theta_B,\psi_B)|$, we can set 
\begin{equation}
    \phi_B = \frac{1}{3} \mathrm{arg}\left[-\frac{C(\theta_A,\psi_A)}{C(\theta_B,\psi_B)} \right] \quad .
\end{equation}
There are 3  remaining constraints that give the geometric prefactors for the three $m=0$ terms. With 4 constraints and 4 variables, we used a Mathematica routine to find numerical solutions, resulting in a non-unique solution where the anisotropic terms are numerically negligible.  
The resulting set of six polarization and momentum directions are given in Table~\ref{tab:6photons}, and the Euler angles in Table~\ref{tab:6angles}.
\begin{table}[]
    \centering
    \begin{tabular}{c|c c c }
        & $\hat{x}$ & $\hat{y}$ & $\hat{z}$ \\
    \hline
     $\hat{\epsilon}_1$ &  $\phantom{-}$0.9367417879781806  & 0 & $-$0.3500211745815407 \\
    $\hat{k}_1$ &   $-0.2756350715435098$ &  0.6163387569543059 & $-$0.7376664856228899 \\
    \hline
     $\hat{\epsilon}_2$ & -0.4683708939890903 & 0.8112421851755609 & -0.3500211745815407 \\
     $\hat{k}_2$ &  -0.3959474850875968 & -0.5468763526077737 & -0.7376664856228899 \\ 
     \hline
     $\hat{\epsilon}_3$ & -0.4683708939890903 & -0.8112421851755609 & -0.3500211745815407 \\
     $\hat{k}_3$ &   0.6715825566311067 & -0.0694624043465322 & -0.7376664856228899\\
     \hline
     $\hat{\epsilon}_4$ & -0.2756350715435098 & 0.6163387569543059 & 0.7376664856228899 \\
     $\hat{k}_4$ & -0.9367417879781806 & 0 & -0.3500211745815407 \\
     \hline
     $\hat{\epsilon}_5$ &  -0.3959474850875968 & -0.5468763526077737 & 0.7376664856228899 \\
     $\hat{k}_5$ &   0.4683708939890903 & -0.8112421851755609 & -0.3500211745815407\\
     \hline
     $\hat{\epsilon}_6$ & 0.6715825566311067 & -0.0694624043465322 & 0.7376664856228899 \\
     $\hat{k}_6$ &  0.4683708939890903 & 0.8112421851755609 & -0.3500211745815407\\
     \hline
        \end{tabular}
    \caption{An optimized solution for 6 sets of polarization and momentum that yield an isotropic x-ray spectra at both dipole and quadrupole.}
    \label{tab:6photons}
\end{table}

\begin{table}
  \centering
    \begin{tabular}{c|c | c | c }
    & $\theta$ & $\phi$ & $\psi$ \\
    \hline
 1 & 1.9283900348468623 & 0 &    
     0.6640848873532189 \\
 2 & 
     1.9283900348468623  & $\frac{2 \pi }{3}$ & 0.6640848873532189 \\
 3 & 1.9283900348468623 & $\frac{4 \pi }{3}$ & 0.6640848873532189 \\ 
 4 & 0.7411887378042126 & 
     1.9913306620788617 & 1.0257906472519933 \\
 5 & 0.7411887378042126 & 4.0857257644720572 & 1.0257906472519933 \\
 6 & 0.7411887378042126 &    
     6.1801208668652527 & 1.0257906472519933  
   \end{tabular}
    \caption{An optimized solution for 6 sets of Euler angles in radians that yield an isotropic x-ray spectra at both dipole and quadrupole order.}
    \label{tab:6angles}
\end{table}

\newpage
\subsection{Special cases}
For reader's interest, below we also provide the solutions of the Euler angles $(\theta, \phi, \psi)$ and weights of the seven special cases defined by space groups in Sec.~5.1 in the paper of Brouder~\cite{brouder1990angular}.
\begin{enumerate}
    \item Cubic point groups O$_h$(m3m), T$_d$($\bar{4}$3m), O(432), T$_h$ (m3) and T(23), two photons: $(\frac{\pi}{2}, -\frac{\pi}{4}, \frac{\pi}{2})$ with weight 0.4 and $(0, \frac{\pi}{4}, 0)$ with weight 0.6. 
    
    \item Groups D$_{\infty h}$($\infty$/mm), C$_{\infty v}$($\infty$m), D$_{6h}$(6/mmm), D$_{3h}$($\bar{6}$m2), C$_{6v}$(6mm),
D$_6$(622), C$_{6h}$(6/m), C$_{3h}$($\bar{6}$) and C$_6$(6), two photons: $(0, 0, 0)$ with weight $\frac{1}{15}$ and $ \left (\arccos \left (\sqrt{\frac{1}{7} } \right ), 0, \arcsin \left ( \sqrt{\frac{5}{12}} \right ) \right )$ with weight $\frac{14}{15}$.
    
    \item Groups D$_{3d}$($\bar{3}$m) and D$_3$(32), two photons: $(0, 0, 0)$ with weight $\frac{1}{15}$ and \\
    $ \left (\arccos \left ( \sqrt{\frac{1}{7}} \right ), \frac{1}{3}\arctan \left (\frac{\sqrt{5}}{2} \right ), \arcsin \left ( \sqrt{\frac{5}{12}} \right )  \right )$ with weight $\frac{14}{15}$.
    
    \item Group C$_{3v}(3m)$, two photons: $(0, 0, 0)$ with weight $\frac{1}{15}$ and \\
    $ \left (\arccos \left ( \sqrt{\frac{1}{7}} \right ), -\frac{1}{3}\arctan \left (\frac{2}{\sqrt{5}} \right ), \arcsin \left ( \sqrt{\frac{5}{12}} \right ) \right )$ with weight $\frac{14}{15}$.
    
    \item Groups S$_6$($\bar{3}$) and C$_3$(3), four photons: $(0,0,0)$ with weight $\frac{1}{15}$, \\
    $ \left (\arccos \left ( \sqrt{\frac{1}{7}} \right ), \frac{1}{3}\arctan \left (\frac{\sqrt{5}}{2} \right ), \arcsin \left ( \sqrt{\frac{5}{12}} \right ) \right )$ with weight $\frac{14}{15}(0.5-\frac{\sqrt{5}}{10})$, \\
    $ \left (\arccos \left ( \sqrt{\frac{1}{7}} \right ), -\frac{1}{3}\arctan \left (\frac{2}{\sqrt{5}} \right ), \arcsin \left ( \sqrt{\frac{5}{12}} \right ) \right )$ with weight $\frac{14}{15}(-0.25+\frac{\sqrt{5}}{4})$, and \\
    $ \left (\arccos \left ( \sqrt{\frac{1}{7}} \right ),  \pi, \arcsin \left ( \sqrt{\frac{5}{12}} \right )  \right )$ with weight $\frac{14}{15}(0.75-\frac{3\sqrt{5}}{20})$.
    
    \item Groups D$_{4h}$(4/mmm), D$_{2d}$($\bar{4}$2m), C$_{4v}$(4mm), and D$_4$(222), two photons: $(0, 0, 0)$ with weight $\frac{1}{15}$ and 
    $ \left (\arccos \left (\sqrt{\frac{1}{7} } \right ), -\frac{1}{4}\arctan \left (\frac{2}{\sqrt{5}} \right ), \arcsin \left ( \sqrt{\frac{5}{12}} \right ) \right )$ with weight $\frac{14}{15}$.
    
    \item Groups C$_{4h}$(4/m), S$_4$($\bar{4}$) and C$_4$(4), four photons: $(0, 0, 0)$ with weight $\frac{1}{15}$, \\
    $ \left (\arccos \left (\sqrt{\frac{1}{7} } \right ), -\frac{1}{4}\arctan \left (\frac{2}{\sqrt{5}} \right ), \arcsin \left ( \sqrt{\frac{5}{12}} \right ) \right )$ with weight $\frac{14}{15} \cdot \frac{\sqrt{5}}{3}/(\frac{\sqrt{5}}{3}+\frac{5}{3})$, \\
    $ \left (\arccos \left (\sqrt{\frac{1}{7} } \right ), \frac{1}{4}\arctan \left (\frac{\sqrt{5}}{2} \right ), \arcsin \left ( \sqrt{\frac{5}{12}} \right ) \right )$ with weight $\frac{14}{15} \cdot \frac{2}{3}/(\frac{\sqrt{5}}{3}+\frac{5}{3})$, and \\
    $ \left (\arccos \left ( \sqrt{\frac{1}{7}} \right ),  \frac{\pi}{4}, \arcsin \left ( \sqrt{\frac{5}{12}} \right )  \right )$ with weight $\frac{14}{15}/(\frac{\sqrt{5}}{3}+\frac{5}{3})$.
\end{enumerate}

\bibliographystyle{unsrt}
\bibliography{references}

\ifdefined\INCLUDED
\section*{Supporting Information}
\else
\documentclass[11pt,aps,prb,amsmath,amssymb,superscriptaddress,showpacs,floatfix,longbibliography]{revtex4-2}
\usepackage{graphicx} 
\usepackage{multirow}
\usepackage[table,xcdraw]{xcolor}
\usepackage{amsmath}
\newcommand{\code}[1]{\texttt{#1}}

\begin{document}
\title{Supporting Information: Benchmark of First-Principles Titanium K-Edge X-Ray Absorption Spectral Simulations on Titanium-containing Oxides}
\author{Chuntian Cao}
\affiliation{Computing and Data Sciences Directorate, Brookhaven National Laboratory, Upton, New York 11973, USA
}%

\author{Joshua J. Kas}
\affiliation{Department of Physics, University of Washington, Seattle, Washington 98195, United States
}%

\author{Bruce Ravel}
\affiliation{Material Measurement Laboratory, National Institute of Standards and Technology, Gaithersburg, Maryland 20899, United States
}%

\author{John Vinson}
\affiliation{Material Measurement Laboratory, National Institute of Standards and Technology, Gaithersburg, Maryland 20899, United States
}%

\author{Deyu Lu}
\affiliation{ 
Center for Functional Nanomaterials,
Brookhaven National Laboratory, Upton, New York 11973,
United States}

\date{\today}
\maketitle

\fi
\renewcommand{\thefigure}{S\arabic{figure}}
\renewcommand{\thetable}{S\arabic{table}}
\renewcommand{\theequation}{S\arabic{equation}}
\setcounter{figure}{0}



\begin{figure}[htbp]
    \centering
    \includegraphics[trim = {0 0  236 0},clip=true]{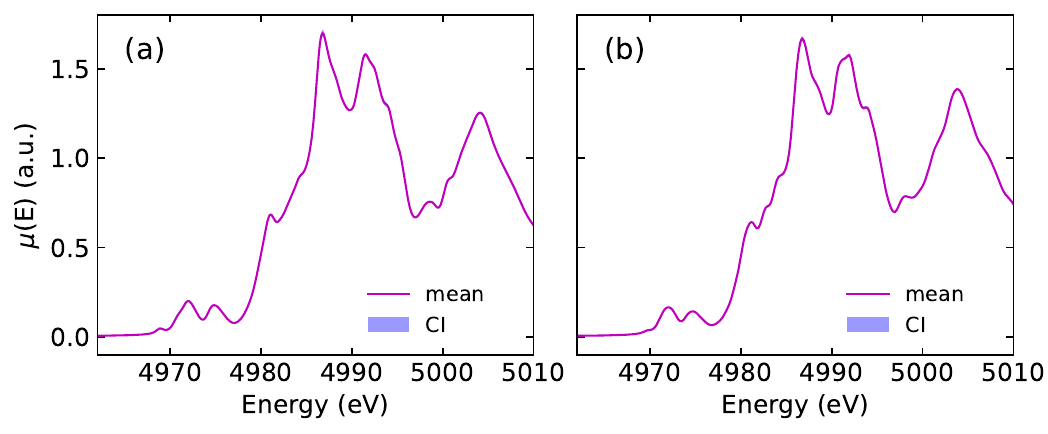}
    \caption{Convergence of thermal average of (a) {\sc ocean} and (b) VASP spectra using five AIMD snapshots. In both cases, the confidence interval (CI) is small enough to be occluded by the linewidth of the mean.  
    }
    \label{fig:SIconv}
\end{figure}

 \vspace{5mm}
    \begin{figure}[htbp]
        \centering
        \includegraphics[width=6.5 in]{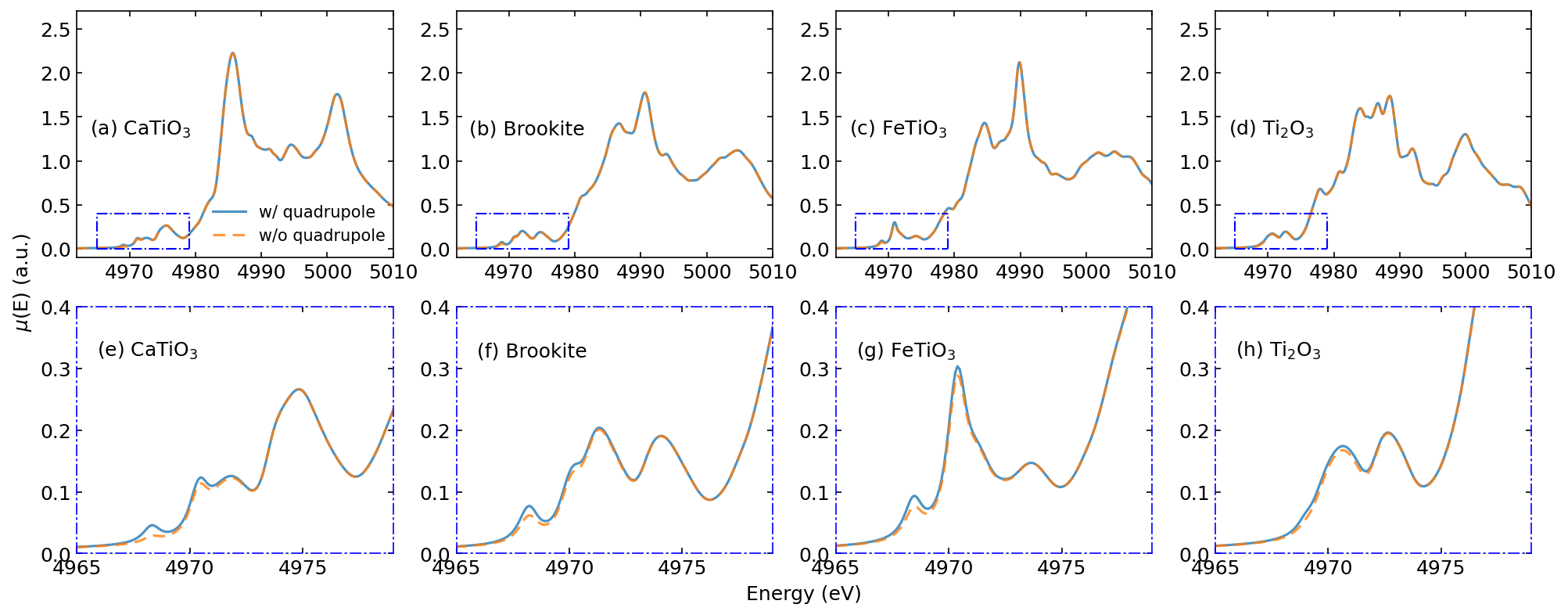}
        \vspace{-0mm}
        \caption{
        Quadrupole contribution to the pre-edge region of Ti K-edge {\sc ocean} spectra. 
        (a--d) Full range XANES spectra. 
        (e--h) Magnified view of pre-edge region. 
        (a,e) CaTiO$_3$, (b,f) Brookite, (c,g) FeTiO$_3$, (d,h) Ti$_2$O$_3$. 
        In each subplot, the blue solid line and yellow dashed line indicate simulations with and without quadrupole contributions, respectively. 
        The spectra are thermally averaged spectra calculated from AIMD snapshots using {\sc ocean} code. Many-body effects are not incorporated. 0.89 eV Lorentzian broadening is used.         
        }
        \label{fig:SI_quad_correction}
    \vspace{2mm}
    \end{figure}    

 \vspace{5mm}
    \begin{figure}[htbp]
        \centering
        \includegraphics[width=6.5 in]{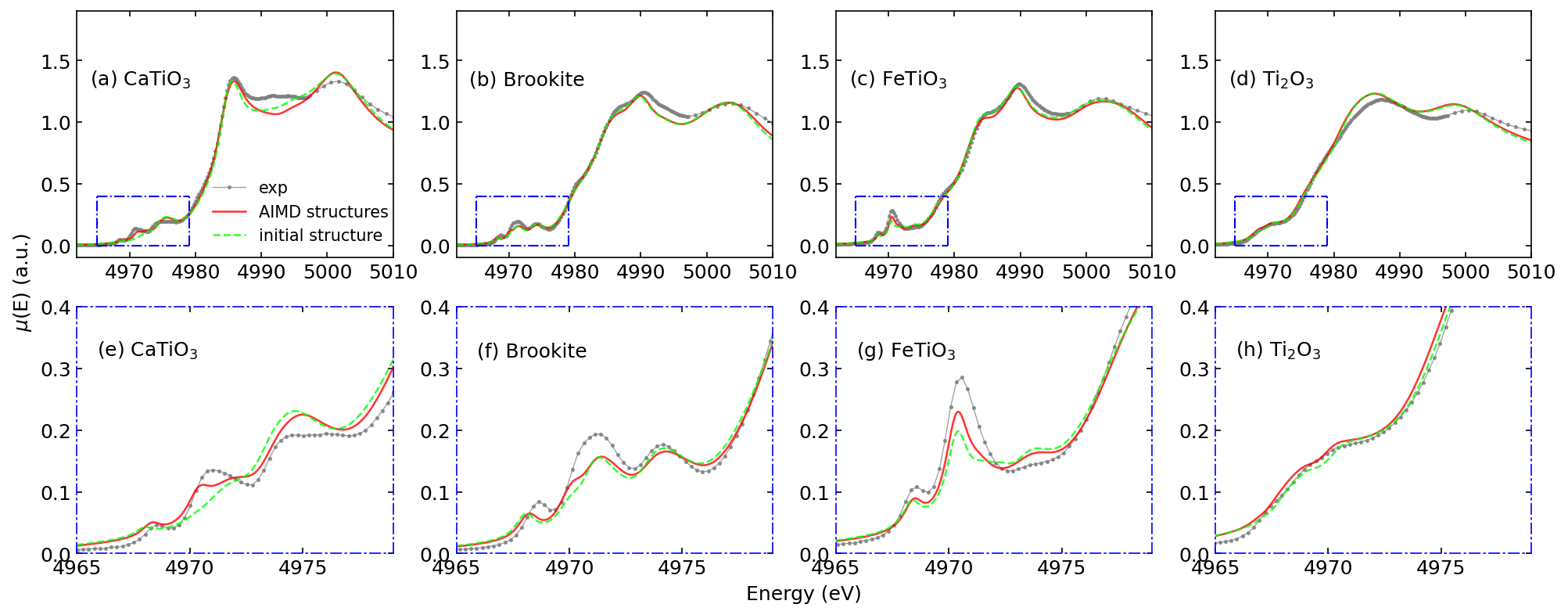}
        \vspace{-0mm}
        \caption{
        Effect of thermal disorder on Ti K-edge spectra based on {\sc ocean} simulations.
        (a--d) Full range XANES spectra. 
        (e--h) Magnified view of pre-edge region; the simulated spectra are aligned to the experimental pre-edge region to better visualize the agreement. 
        (a,e) CaTiO$_3$, (b,f) Brookite, (c,g) FeTiO$_3$, (d,h) Ti$_2$O$_3$. 
        In each subplot, grey lines and markers represent experimental spectra, red solid lines represent spectra thermally averaged from simulated spectra from AIMD snapshots, and green dashed lines represent simulated spectra from static experimental structure without thermal disorder. 
        All simulated spectra include quadrupole contributions and many-body shake-up effects, and energy-dependent Lorentzian-broadening. 
        }
        \label{fig:SI_thermal_correction}
    \vspace{2mm}
    \end{figure}    

 \vspace{5mm}
    \begin{figure}[htbp]
        \centering
        \includegraphics[width=6.5 in]{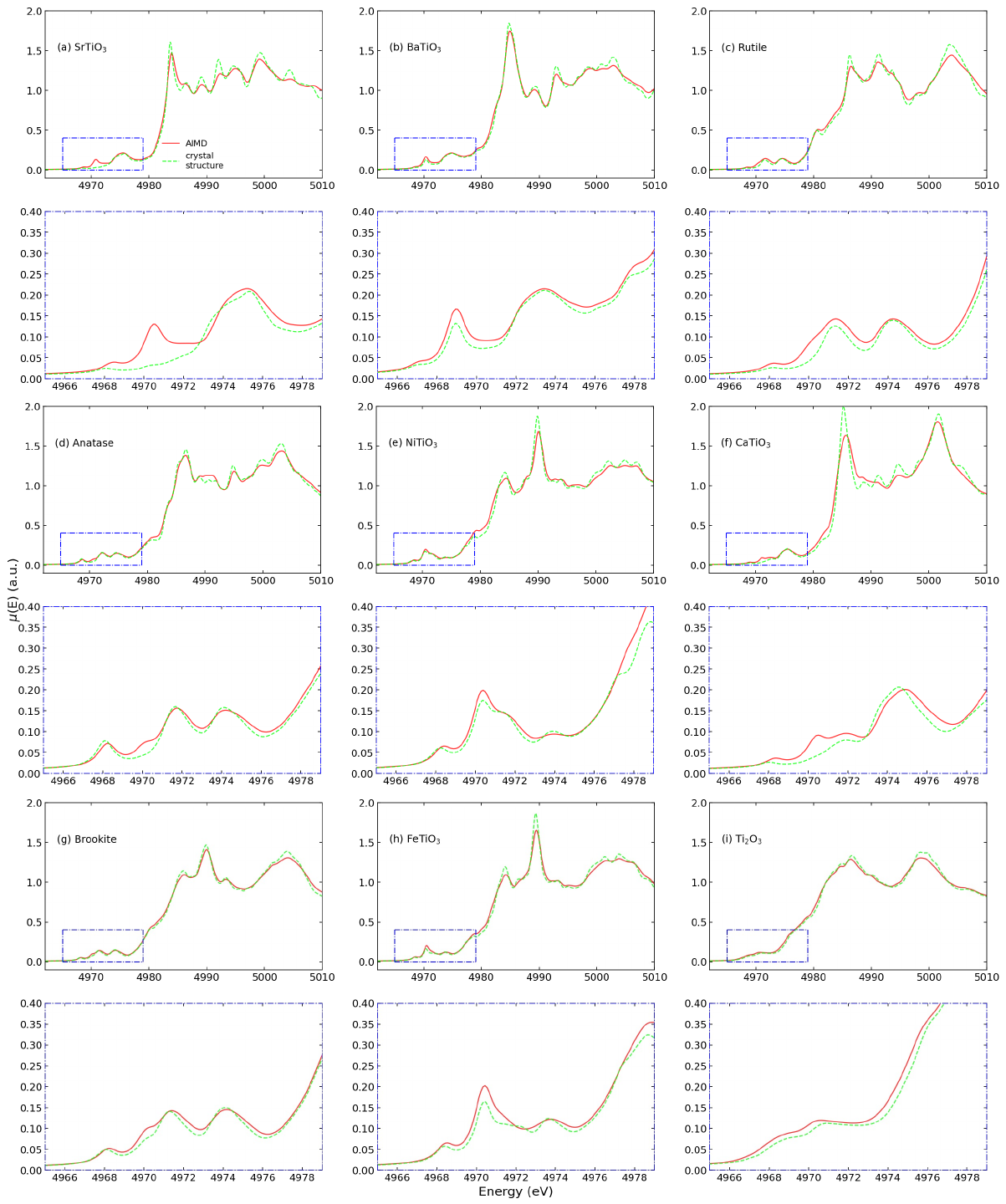}
        \vspace{-0mm}
        \caption{
        Effect of thermal disorder on Ti K-edge spectra of the full energy range and the pre-edge region based on {\sc ocean} simulations.
        In each subplot, red solid lines represent spectra thermally averaged from simulated spectra from AIMD snapshots, and green dashed lines represent simulated spectra from static experimental structure without thermal disorder. 
        All simulated spectra include quadrupole contributions and many-body shake-up. 
        A 0.89 eV Lorentzian broadening rather than energy-dependent Lorentzian-broadening is applied. 
        }
        \label{fig:SI_thermal_noBroaden}
    \vspace{2mm}
    \end{figure}    

 \vspace{5mm}
    \begin{figure}[htbp]
        \centering
        \includegraphics[width=6.5 in]{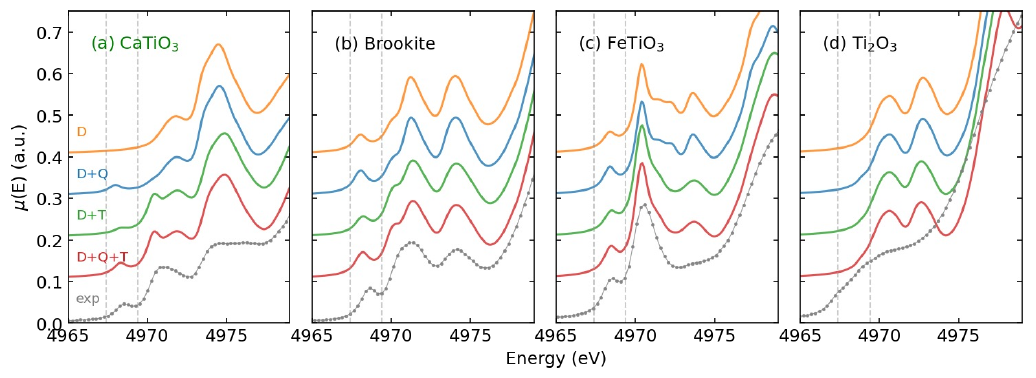}
        \vspace{-0mm}
        \caption{
        Pre-edge regions showing the effect of thermal disorder and quadrupole contributions based on {\sc ocean} simulations.
        Orange: dipole; blue: dipole plus quadrupole; green: dipole plus thermal disorder; red: dipole plus quadrupole plus thermal disorder; grey markers and lines: experimental spectra.  Simulations do not include many-body shake-up and energy-dependent broadening in order to isolate pre-edge effects.
        }
        \label{fig:SI_pre_edge_noBroaden}
    \vspace{2mm}
    \end{figure}    

 \vspace{5mm}
    \begin{figure}[htbp]
        \centering
        \includegraphics[width=\textwidth]{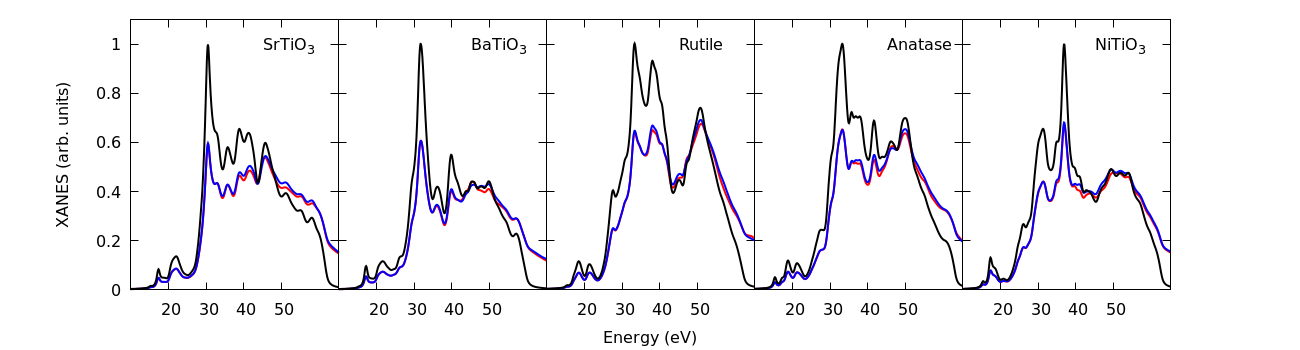}
        \vspace{-0mm}
        \caption{
        Comparison of the {\sc ocean} spectra (black) with that convolved with the full many-body spectral function (red) or with the model spectral function (blue). 
        }
        \label{fig:SI_xas_conv}
    \vspace{2mm}
    \end{figure}   

 \vspace{5mm}
    \begin{figure}[htbp]
        \centering
        \includegraphics[width=6.5 in]{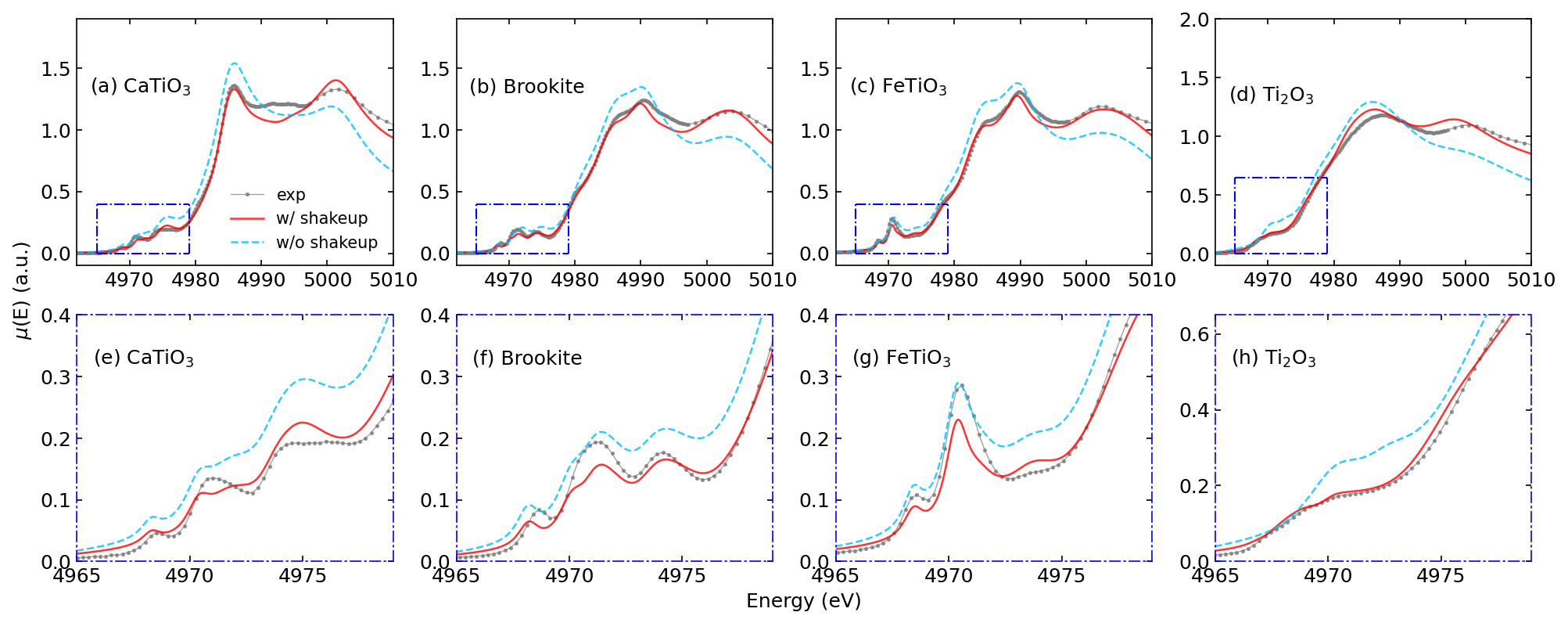}
        \vspace{-0mm}
        \caption{
        Effect of many-body shakeup on Ti K-edge spectra based on {\sc ocean} simulations.
        (a--d) Full range XANES spectra. 
        (e--h) Magnified view of pre-edge region; the simulated spectra are aligned to the experimental pre-edge region to better visualize the agreement. 
        (a,e) CaTiO$_3$, (b,f) Brookite, (c,g) FeTiO$_3$, (d,h) Ti$_2$O$_3$. 
        In each subplot, grey lines and markers represent experimental spectra, red solid lines represent simulated spectra with many-body shake-up, and blue dashed lines represent simulated spectra without many-body shake-up. 
        All simulations include quadrupole contributions and thermal disorder, as well as energy-dependent Lorentzian-broadening. 
        }
        \label{fig:SI_shakeup_correction}
    \vspace{2mm}
    \end{figure}    

 \vspace{5mm}
    \begin{figure}[htbp]
        \centering
        \includegraphics[width=\textwidth]{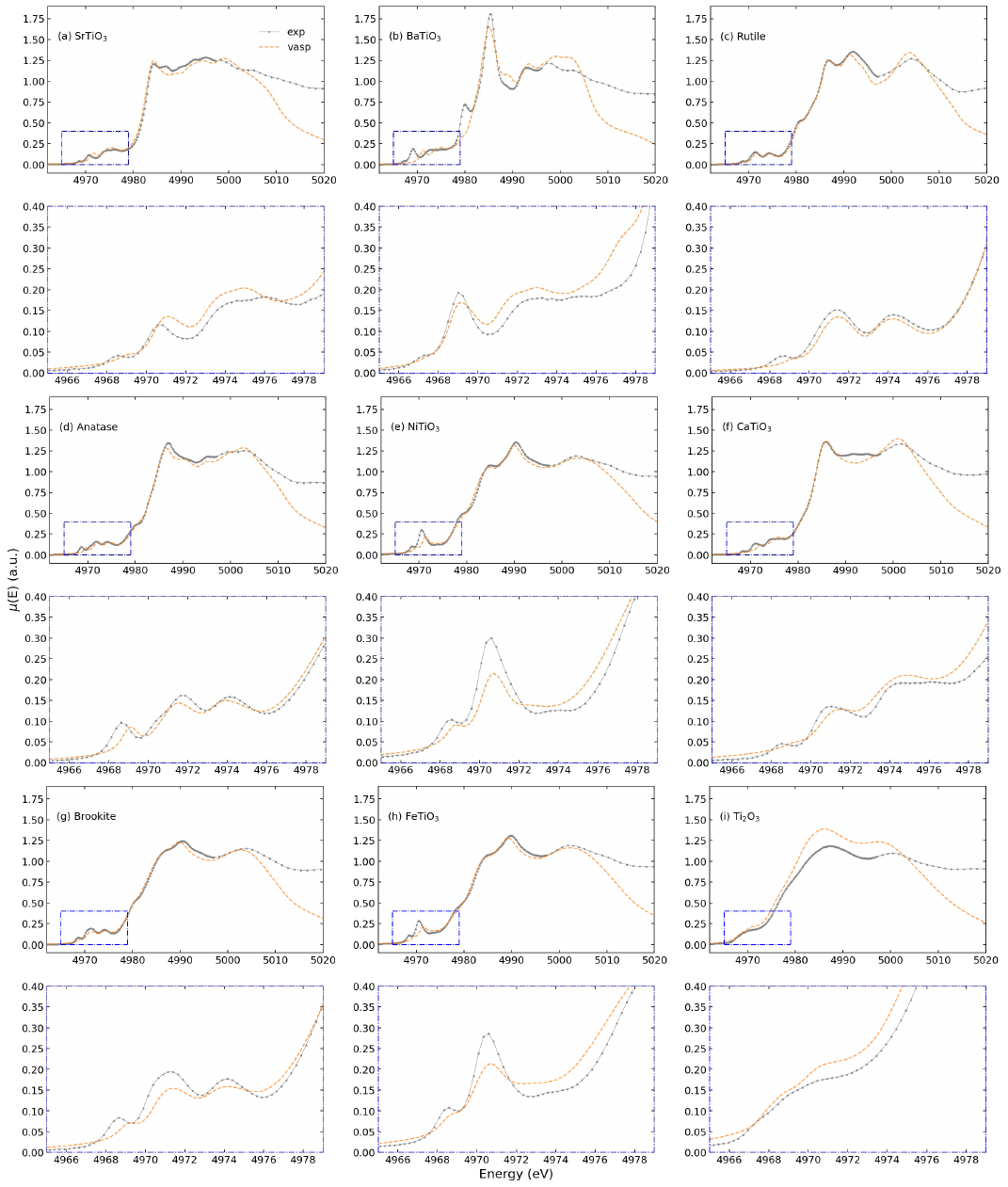}
        \vspace{-0mm}
        \caption{
        Comparison of VASP spectra (orange) with experiment (black) in the full energy range and the pre-edge region. 
        }
        \label{fig:SI_exp_vasp}
    \vspace{2mm}
    \end{figure}   

\vspace{5mm}
    \begin{figure}[htbp]
        \centering
        \includegraphics[width=4.5 in]{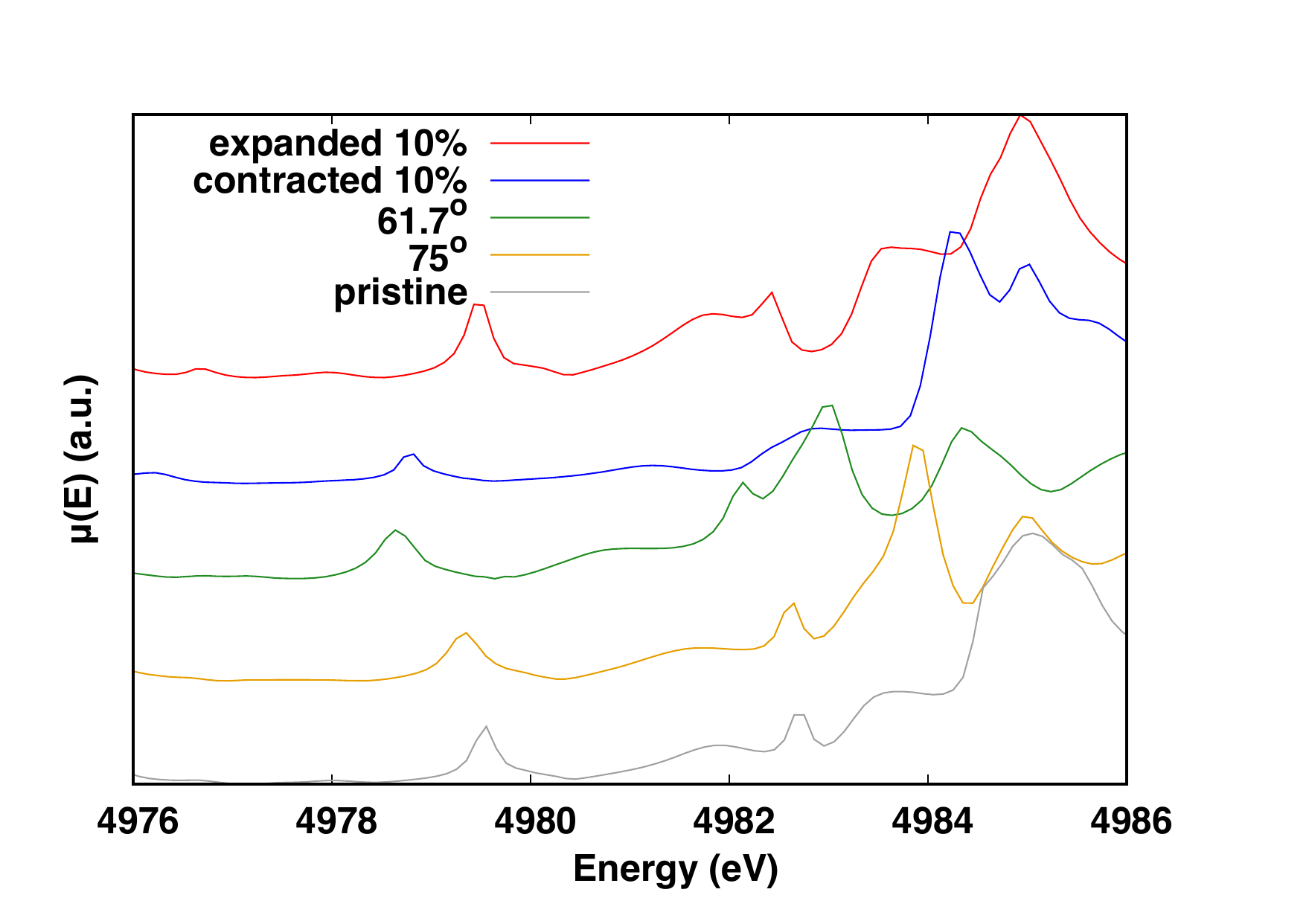}
        \vspace{-0mm}
        \caption{FEFF10 calculations of the XANES showing the effect of Jahn-Teller like distortion as well as equatorial bond angle distortion. The curves were calculated by expanding/contracting the apical O bonds by 10$\%$ (red/blue) and reducing the equatorial O-Ti-O bond angle from $89.7^o$ to $75^o$ (green) and $61.7^o$ (orange). The results of the pristine calculation are also shown in gray.
        }
        \label{fig:SI-feff_distortion}
    \vspace{2mm}
    \end{figure}   

\vspace{5mm}
    \begin{figure}[htbp]
        \centering
        \includegraphics[width=4 in]{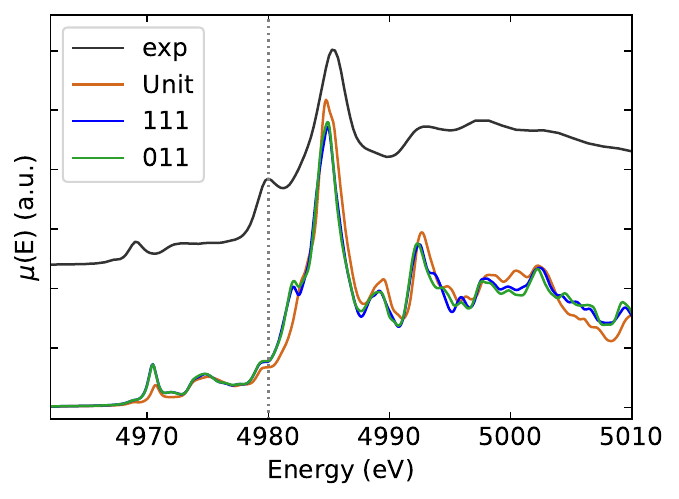}
        \vspace{-0mm}
        \caption{The effect of distortion on the Ti K-edge spectra of BaTiO$_3$ compared to the undistorted unit cell calculated with {\sc ocean}. While the main edge peak develops a new shoulder, the discrepancy with experiment remains. 
        }
        \label{fig:SI-ocean-metastable}
    \vspace{2mm}
    \end{figure}

\begin{figure}[tbph!]
    \centering
    \includegraphics[width=3.0 in]{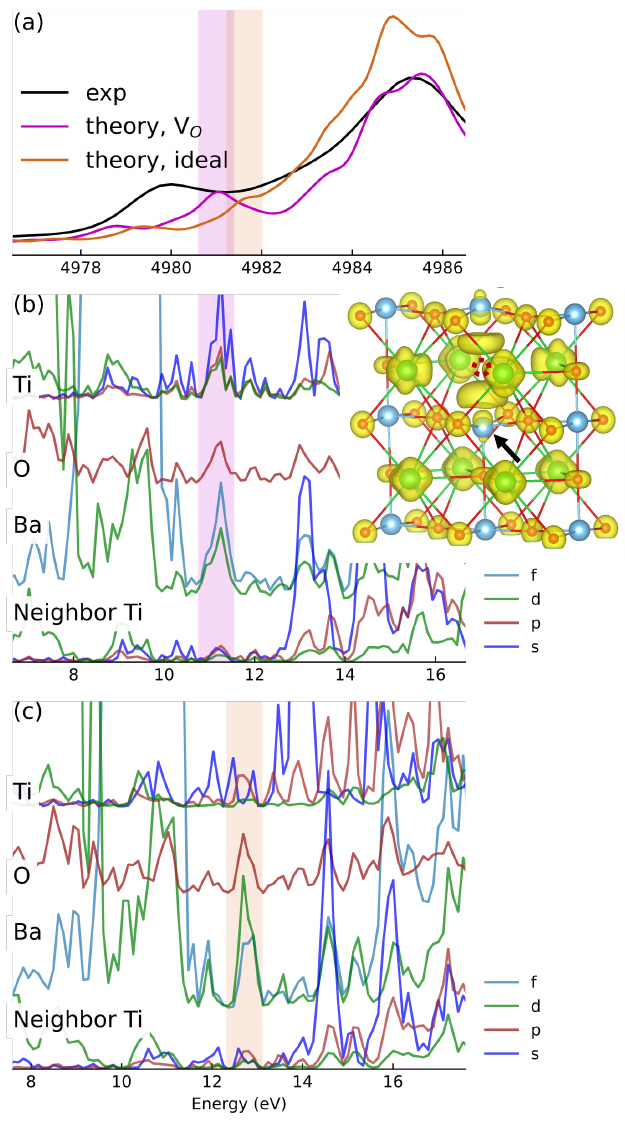}
    \caption{(a) Comparison of the measured spectrum with the simulated spectra of the five-fold Ti site next to the oxygen vacancy (V$_\text{O}$) and the ideal octahedral structure. (b) PDOS of the $3\times 3\times 3$ supercell with an oxygen vacancy (dashed red circle in the structure plot) in the presence of a full core hole on the Ti absorber atom (indicated by the arrow in the structure plot). The emerged shoulder is highlighted in purple. Inset shows the isosurface of the partial charge density corresponding to the shoulder. (c) PDOS of the ideal structure. The small shoulder is highlighted in orange.}
    \label{fig:pdos-isosurface}
\end{figure}

\if{0}
    \begin{figure}[htbp]
        \centering
        \includegraphics[width=4 in]{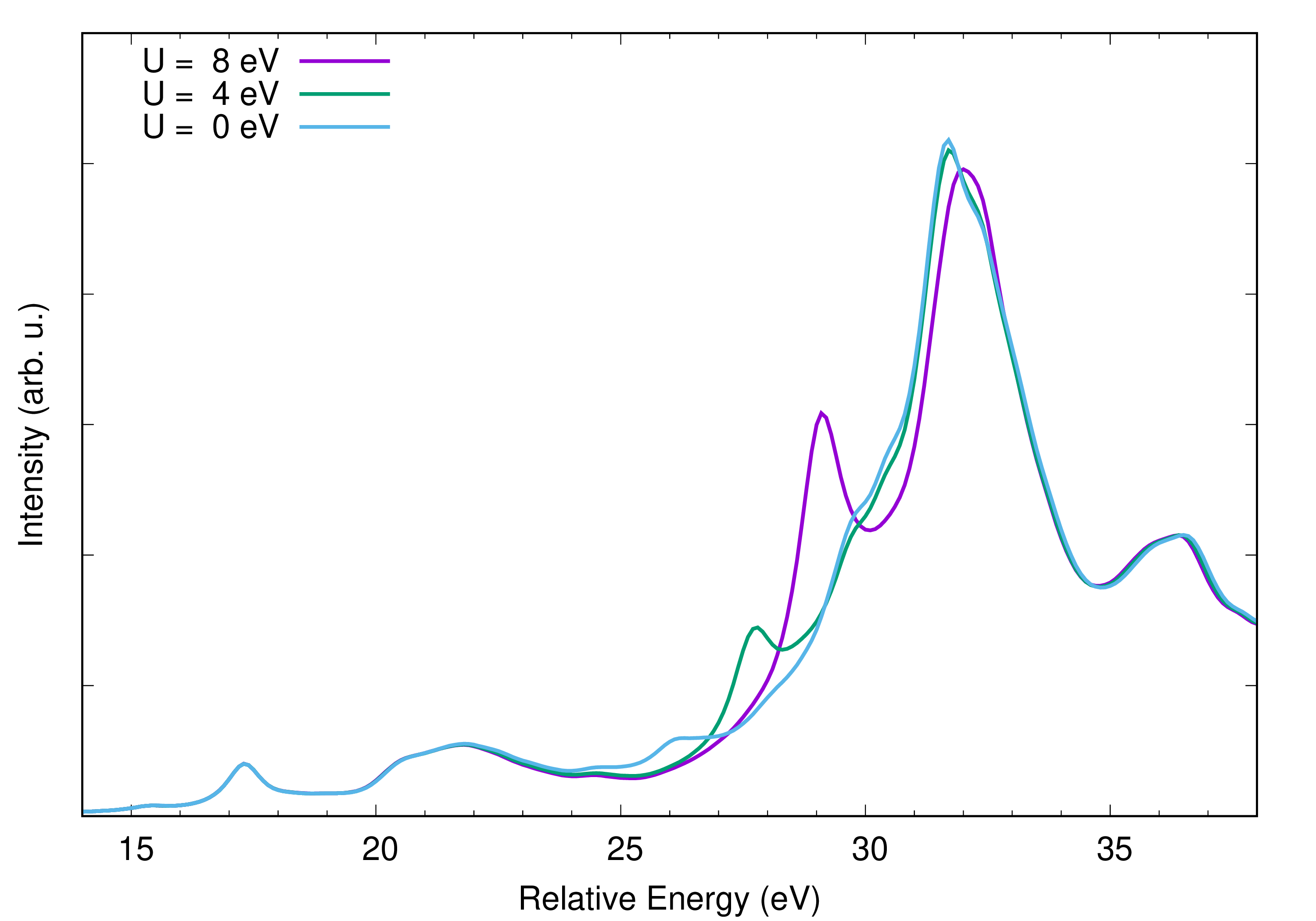}
        \vspace{-0mm}
        \caption{The effect of a Hubbard U correction on the Ba 4{\it f} states on the Ti K-edge spectra of BaTiO$_3$ as calculated with {\sc ocean}. 
        }
        \label{fig:SI_BTO+U}
    \vspace{2mm}
    \end{figure}   
\fi

\ifdefined\INCLUDED\else
\end{document}
\fi

\end{document}